\documentclass[journal]{IEEEtran}

\usepackage{psfrag}
\usepackage{amsmath}
\usepackage{subfigure}
\usepackage{url}
\usepackage{stfloats}
\usepackage{cite}
\usepackage{amsmath}
\usepackage{mathtools}
\usepackage{upgreek}
\usepackage{amssymb}
\usepackage{amsfonts}
\usepackage{graphicx}
\usepackage{fancybox}
\usepackage{color}
\usepackage{algorithm}
\usepackage{algorithmic}
\usepackage{multirow}
\usepackage{booktabs}
\usepackage{threeparttable}
\usepackage{latexsym}
\usepackage{bm,array}
\usepackage{graphicx}
\usepackage{float}
\usepackage{hyperref}
\usepackage{acronym}
\usepackage{textcomp}

\IEEEoverridecommandlockouts
\normalsize
\begin{document}

\title{Cellular Communications in Ocean Waves for Maritime Internet of Things
\\
}
\author{
Yiming~Huo,~\IEEEmembership{Member,~IEEE},
Xiaodai~Dong,~\IEEEmembership{Senior Member,~IEEE},
and Scott~Beatty
\\
\thanks{Y. Huo and X. Dong are with the Department of Electrical and Computer Engineering, University of Victoria, Victoria, BC V8P 5C2, Canada (e-mail: ymhuo@uvic.ca, xdong@ece.uvic.ca).} 
 \thanks{S. Beatty 
 is with MarineLabs Data Systems Inc., 2A-4476 Markham St., Victoria, BC, Canada (e-mail: scott@marinelabs.io).} 
 \thanks{This work was supported by the NSERC of Canada. \emph{(Corresponding author: Xiaodai Dong)}}
}





\maketitle
\begin{abstract}
The rapid advancement of Internet of Things (IoT) and 5G and beyond technologies are transforming the marine industry and research. Our understanding of the vast sea that covers 71\% of the Earth's surface is being enhanced by the various ocean sensor networks equipped with effective communications technologies. In this paper, we begin with a review of the research and development status-quo of maritime IoT (MIoT) enabled by multiple wireless communication technologies. Then we study the impact of sea waves to radio propagation and the communications link quality. Due to the severe attenuation of sea water to radio frequency electromagnetic waves propagation, large ocean waves can easily block the communications link between a buoy sensor and a cell tower near shore. This paper for the first time uses the ocean wave modeling of coastal and oceanic waters to examine the line of sight communications condition. Real wave measurement data parameters are applied in the numerical evaluation of the developed model. Finally, the critical antenna design taking into account the wave impact is numerically studied with implementation solutions proposed, and the system hardware and protocol aspects are discussed.         
\end{abstract}

\begin{IEEEkeywords}
Internet of Things (IoT), Maritime IoT (MIoT), machine-type communication (MTC), ocean engineering, ocean wave characterization, near-shore communication, wireless sensor networks, terrestrial-oceanic hybrid communication, antenna.   
\end{IEEEkeywords}

\section{Introduction}
The fifth generation (5G) wireless communication and Inter-net of Things (IoT) technologies are fundamentally reshaping human societies. The underlying technologies of 5G and IoT such as real-time computing, machine learning, signal processing, wireless communications, big data, etc., have been advancing significantly, which enables many promising new applications. In light of the new research and development trend in cellular communications, the applications are not only limited to conventional terrestrial communications, but also quickly expanding to aerial and outer-space domains. \textcolor{black}{For example, the availability of low-cost high-performance unmanned aerial vehicles (UAVs) has resulted in the feasibility of deploying UAV base stations \cite{Hossein:UAV}, \cite{Huo:VTC}, cellular-connected UAV user equipment (UE) \cite{Bergh:UAV}, \cite{Al:IoT}, UAV-assisted IoT networks \cite{Huo:DAMU}, and UAV-assisted mobile edge computing \cite{Li:UAV}. Furthermore, with satellite communications considered as a critical vertical component of the 5G and beyond ecosystem \cite{Huang:Green}, another dimension of IoT technologies is enabled by such as SpaceX’s Starlink which aims to launch tens of thousands of satellites until 2025 \cite{SpaceX:FCC}. Therefore, service offloading and resource management for terrestrial-satellite systems are particularly critical \cite{Gao:Satellite}, \cite{Fu:Satellite}.} 

On the other hand, the concept of maritime IoT (MIoT) developed by United Nations chartered International Maritime Organization (IMO) \cite{Xia:Satellite} has targeted delivering multiple tasks such as ubiquitous connectivity for maritime devices on a global scale, the enhancement of related services for safety and security at sea, marine environment protection, and ocean engineering research. For example, the maritime industry is heading toward autonomous shipping, which requires maritime machine-type communications (MTC) as one of the key enablers. 
In particular, for maritime MTC, automatic identification system (AIS) and application specific message (ASM) system for ships have been developed by International Association of Marine Aids to Navigation and Light-house Authorities (IALA) \cite{IALA:IALA}. More recently, IALA, IMO and other national maritime authorities have started the work on the very high frequency (VHF) Data Exchange System (VDES) to enable the data link supported services such as Vessel Traffic Service (VTS). At World Radio Conference 2015 (WRC-2015), the International Telecommunication Union (ITU) has assigned 100 kHz spectrum for the downlink and uplink of terrestrial VDE part, respectively, and two 25 kHz channels for ASM. 

\begin{figure*}[t]
\begin{center}
\includegraphics[width=7in]{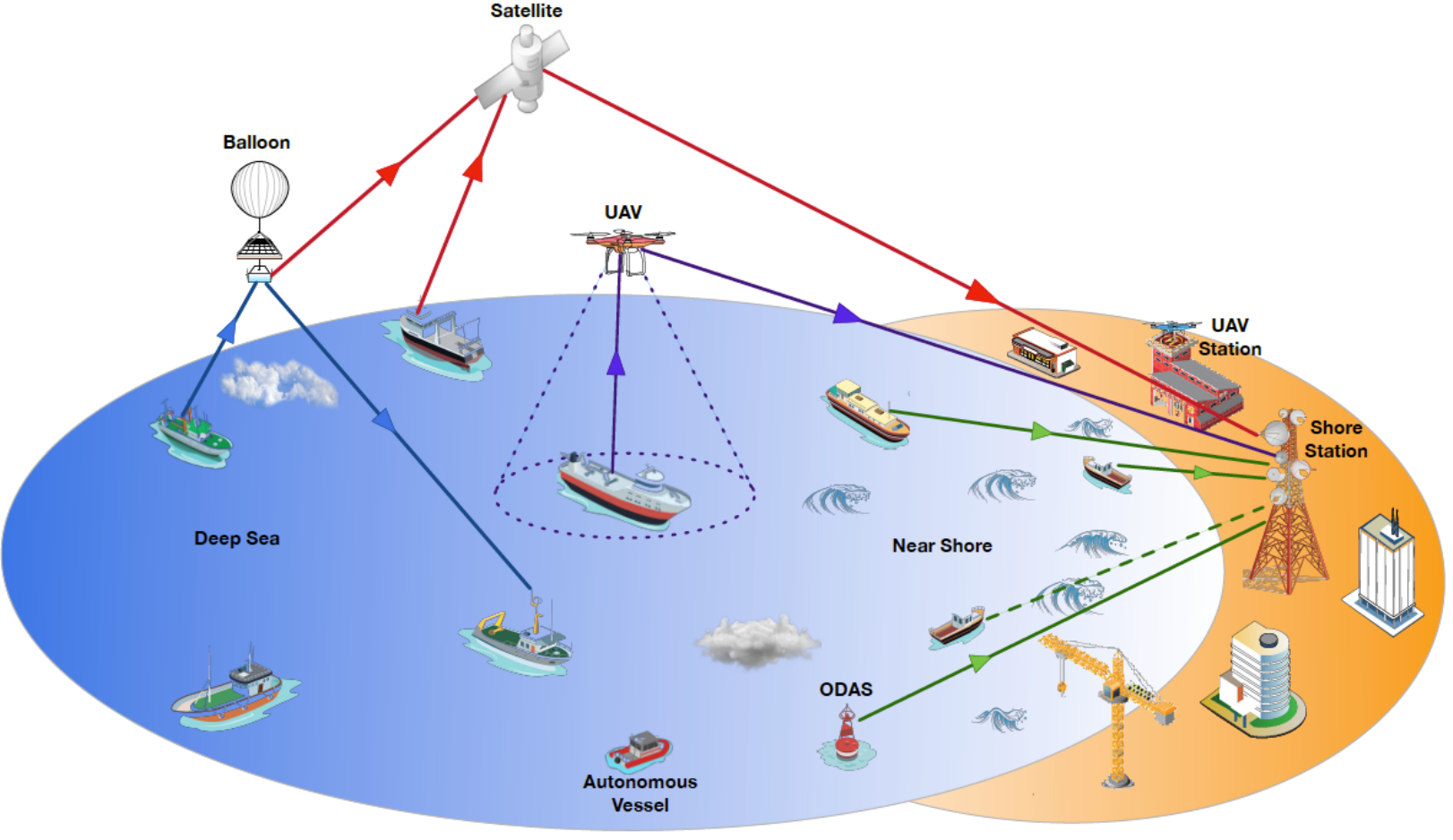}
\caption{Maritime MTC (IoT) communications enabled by various wireless communications technologies.}\label{Fig:MIoT}
\end{center}
\end{figure*}

As shown in Fig.~\ref{Fig:MIoT}, the future maritime MTC includes near-shore communications, high seas communications, and transoceanic communications. In particular, the latter two types may require reliable long-range and higher data rate wireless technologies and networks, e.g., satellites and UAVs. Satellites assist in achieving a global coverage for maritime communications, and the UAV BSs/relays can be deployed to enable a longer transmission distance (e.g., in high/deep sea). Balloon UAVs that are usually deployed at stratospheric layer can enable wider coverage/relay than rotary/fixed-wing UAVs, or co-work with satellites to form more reliable maritime networks particularly when satellite communications are degraded by adverse weather conditions. Although modern satellite and UAV communications have the potential to extend the coverage and enhance the data rate of maritime MTC applications such as autonomous vehicles, their cost of deployment, maintenance and usage fee slow down their adoptions in many MIoT applications. For near shore applications ($<$10 km), some existing commercially mature and ready technologies, such as LTE, NB-IoT, provide alternative and cost-efficient solutions under suitable conditions and configurations. 

Human activities and climate change are increasingly stressing the ocean ecosystems. Data collection and analysis is vital to our ability to manage and sustain our oceans. Ocean data acquisition systems (ODAS) are instruments deployed at sea to collect meteorological and oceanographic data that provide information on the state of the ocean and the surrounding lower atmosphere. They can be attached to buoys, lighthouses, offshore platforms, and measure wave, air/sea temperature, pressure, wind, water composition, etc. Geostationary orbit (GEO) satellites were the default connectivity methods in many ODAS, but they are costly, bulky, and have low data rate with high power consumption. The advancement of wireless communications technologies enables new generation ODAS the capability of higher sensor data sampling rate, real-time data transmission and longer battery  with lower cost. For example, buoy is a widely used multi-tech integrated near-shore unmanned device which can serve multiple purposes including navigation, rescue, oceanic research, and integrating various sensors with cellular communications modules expands its MIoT application scenarios. 

In \cite{Kaz:Marine}, a set of communication technologies from IoT to WiFi and LTE have been adopted to implement a novel communication system for marine monitoring deployment. \textcolor{black}{In \cite{Jo:LTE}, the experimental results have shown that the Korean LTE-Maritime research project could be a practical ship-to-shore data communication solution that can achieve long coverage around 100 km with the order of Mbps. In \cite{Zhou:TRITON}, a wireless mesh network for high-speed and low-cost maritime communications have been carried out in TRITON project in which worldwide interoperability for microwave access (WiMAX) (IEEE 802.16d) serves as communication technology using GPS to achieve time synchronization. Furthermore, in Singapore, WiMAX technology is used to enable a wireless broadband access for seaport (WISEPORT) with the data rate up to 5 Mbps and the coverage range of 15 km \cite{Yang:Maritime}. In \cite{Teixeira:BLUECOMC}, the BLUECOMC project has delivered a vast sea area coverage with the balloon-based relays assistance.} In \cite{Winston:Iridium}, in order to overcome the line-of-sight (LoS) challenge, U.S. National Oceanic and Atmospheric Administration (NOAA) experts have proposed to move the data collection and software to the buoy to incorporate an Iridium Shortest Burst Data modem for primary telemetry and cellular communication. Moreover, \cite{Helmi:Buoy} and \cite{Vengatesan:Buoy} have demonstrated buoy system designs enabled by Global System for Mobile Communication (GSM) and global positioning system (GPS). The cellular technology designed for terrestrial communications, however, will be heavily impacted by the ocean environments, and hence cellular IoT for marine applications faces new challenges that are not yet well understood. The aforementioned papers mainly focus on system and hardware implementation but have not investigated ocean environment impacts.

The maritime communications is characterized by the unique propagation channels different from the terrestrial environment. The diverse application scenarios entail a wide variety of channels such as ship-to-ship, buoy-to-land, ship-to-land, buoy-to-ship, etc. A comprehensive survey of maritime radio propagation channel modeling is given in~\cite{Jin:Maritime}, where all communications are grouped into two types of channels, air-to-sea and near-sea-surface. As detailed in~\cite{Jin:Maritime}, there is fair amount of literature on the wireless channel models for maritime communications, e.g.,~\cite{Haspert:Multipath} -- \cite{Dinc:Surface2}, but many more studies are still needed due to diverse and complicated sea conditions. For near-sea-surface channels, measurements and analytical studies suggest a channel model of three components, LoS, specular reflection path and diffuse reflection multipaths~\cite{Huang:Multipath} -- \cite{Dinc:Surface2}. The contribution of each component depends on the transmitter antenna height, receiver antenna height, distance between transmitter and receiver, location in the sea, sea surface roughness, etc. Although sea surface roughness is included in some channel models through root mean square sea surface height, there has been no studies on the wave shadowing effect, partially because all the measurement campaigns were done on ships with fairly high antenna elevations and it would be very difficult to conduct experiments in large waves. As buoys will play increasingly important roles in MIoT, wave blocking over buoy antennas is a non-negligible phenomenon that will significantly impact the performance of the buoy communications link. In this paper, we investigate the sea wave blocking of line of sight transmission using a detailed wave model for the first time.     


The primary contribution of this paper lies in several folds. First, sea water attenuation to radio wave propagation is presented using the ITU model.  Second, the mathematical modeling of oceanic and coastal waves is summarized to facilitate the statistical analysis of their impacts on the wireless communications. Third, the statistical behavior of LoS due to sea wave blocking is analyzed for communications between an unmanned buoy and on-land cell tower near the shore. Such study is important to energy efficient protocol design to minimize data retransmission times due to loss of LoS. Energy efficiency is critical to battery powered buoy operations. Four, cellular MIoT system specifications are interpreted into hardware design specifications \textcolor{black}{such as effective antenna heights} to further unveil design considerations and provide verifiable solutions in practical oceanic deployment of buoys. To the best of the authors' knowledge, this is the first paper of its kind to provide in-depth analysis of cellular enabled MIoT in large, dynamic ocean waves.

The rest of the paper is organized as follows. Section II presents maritime radio propagation modeling with emphasis on the sea wave blocking of LoS. Numerical results are presented to characterize the probability of LoS communication and continuous LoS duration statistics. Section III proposes energy efficient system and antenna design, considering the wave effect. Section VI concludes the paper.   

\section{MARITIME RADIO PROPAGATION MODELING }


\begin{figure}

\centering
\subfigure[]{\label{}\includegraphics[scale=.7]{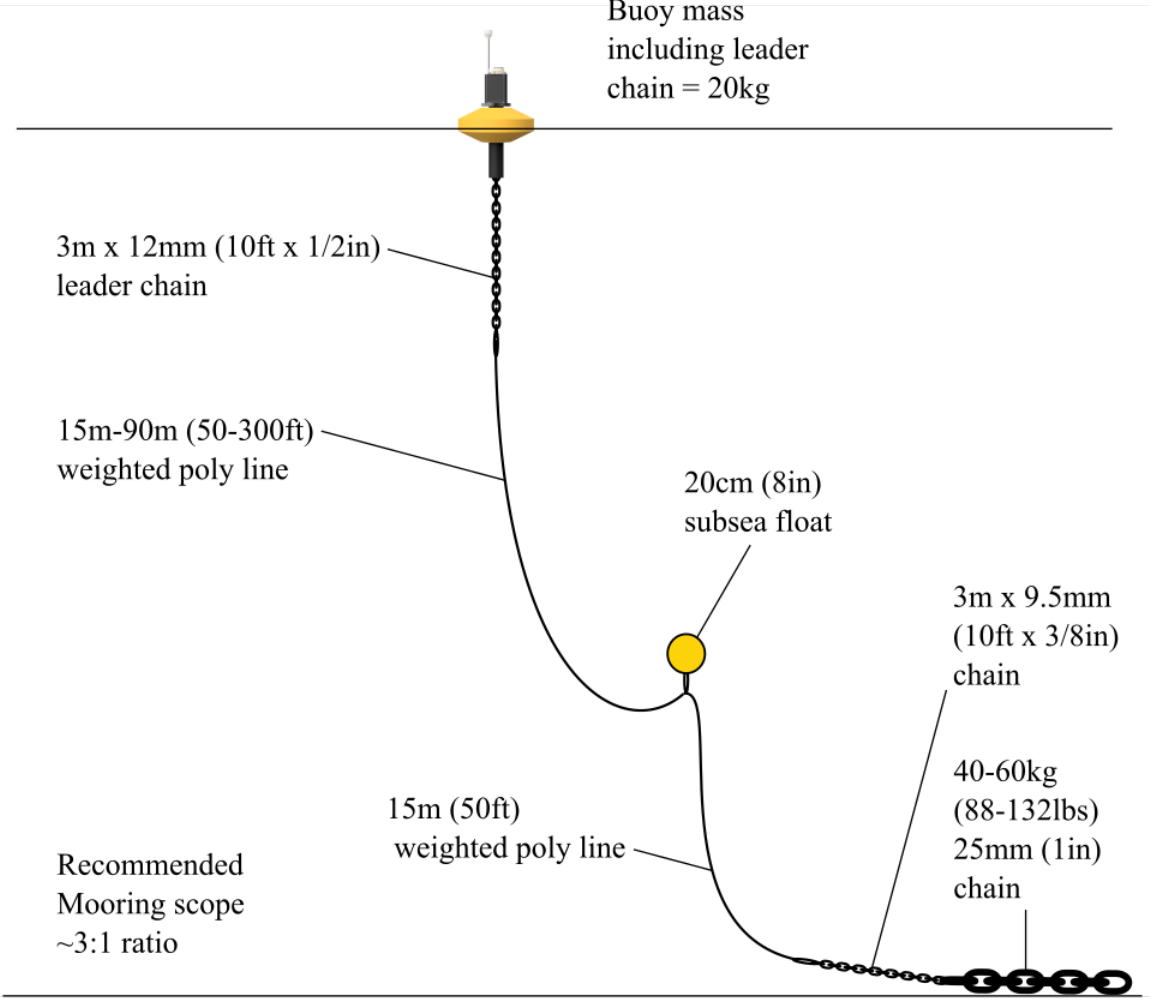}}

\begin{minipage}{.5\linewidth}
\centering
\subfigure[]{\includegraphics[scale=.55]{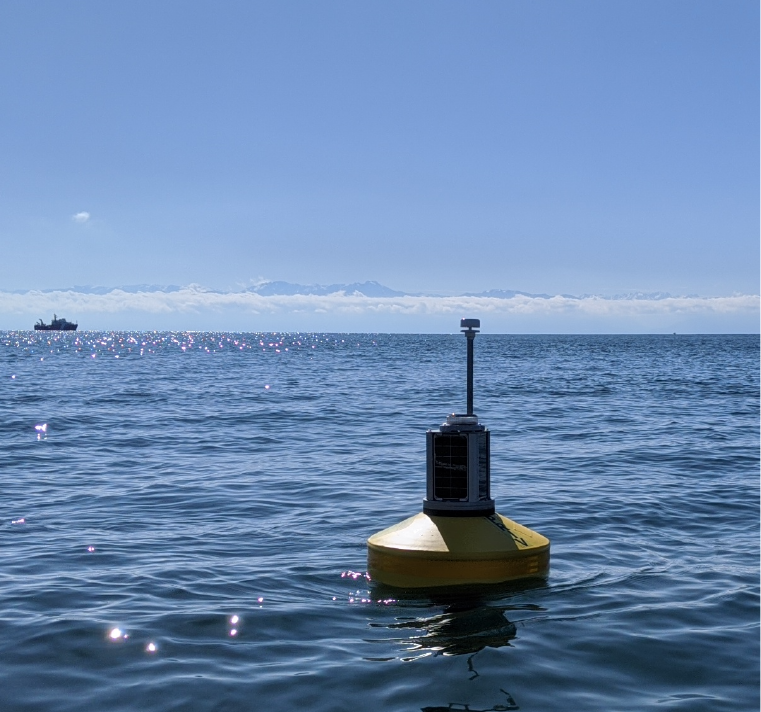}}
\end{minipage}%
\begin{minipage}{.5\linewidth}
\centering
\subfigure[]{\includegraphics[scale=.55]{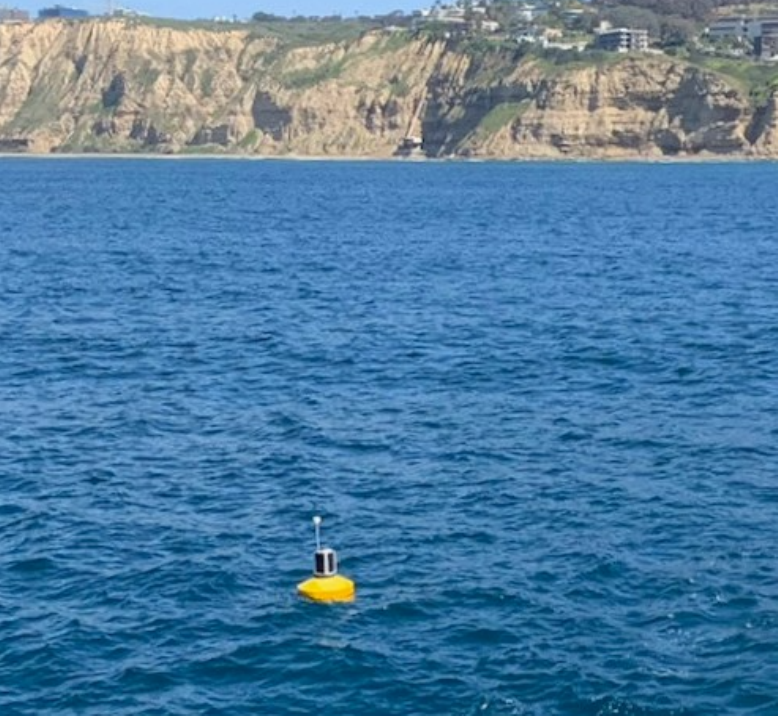}}
\end{minipage}\par\medskip

\caption{\textcolor{black}{(a) Overall dimensions and the typical mooring arrangement for the MarineLabs CoastScout, and CoastScout coastal deployment in the Pacific with (b) calm sea state, and (c) small surface variation.}}\label{fig:Buoy}
\end{figure}

An increasingly popular MIoT device is the new generation compact, rugged and easy-to-deploy buoy, such as MarineLabs' CoastScout \cite{Beatty:Marine}, that measures, records and transmits valuable ocean wave data for a variety of marine applications such as marine safety, port management, ocean engineering, construction, etc. Real-time or quasi real-time transmission of the wave data from these buoys provides significantly higher sampling density of the ocean than in the pre-IoT era and hence a more comprehensive and accurate view of ocean dynamics is obtained. As these buoys are usually deployed near shore (within 2 or 3 km from shoreline), cellular communications between the buoy and terrestrial cellular towers on the shore is a natural choice. Fig.~\ref{fig:Buoy}(a) shows the buoy and mooring system prepared for deployment in the Pacific Ocean in 35-m water depth of Vancouver Island BC \cite{Beatty:Marine}. \textcolor{black}{The buoy (integrated with solar panels) holds a diameter of 60 cm and its mass including a 3 m $\times$ 12 mm leader chain is 15 kg. The weighted poly line has a length range from 15 m to 90 m with a subsea float in the end that joints another 15 m weighted poly line and 3 m $\times$ 9.5 mm chain and 40 -- 60 kg weighted 25 mm chain.} Fig.~\ref{fig:Buoy}(b) and Fig.~\ref{fig:Buoy}(c) illustrate the deployed buoy in the Pacific coastal waters in peaceful and rippled waters, respectively. Data transmission from these buoys to land faces significantly more challenges than terrestrial applications. They operate in harsh ocean environments, power supply is limited, data rate and reliability requirements are high, and rhythmic ocean waves block the radio transmission path regularly.  

A near-sea-surface channel is well known to be modeled as a combination of the LoS path, specular reflection path and diffuse reflection paths due to rough sea surface scattering \cite{Huang:Multipath}. Among them, LoS is the most important link, while the specular path with lower power may add constructively or destructively depending on the path length difference and reflection coefficient. The diffuse paths are always random incoherent components with even smaller total power. A two wave with diffusion power (TWDP) model for terrestrial millimeter wave propagation is considered applicable for the near-sea-surface channel in \cite{Jin:Maritime}. These models work fairly well for calm sea states, as demonstrated by the measurements in \cite{Wang:Baltic} for the 5.2 GHz carrier frequency, transmitter antenna height of 7 m, receiver antenna height of 32.9 m, bandwidth 100 MHz and distance up to 10 km. For rough sea states and/or a low device antenna height, varying sea surface may reach higher than a point on the LoS path. Reflection and scattering in these cases become very difficult to predict and model. In this section, we analyze the wave blocking of LoS phenomenon which involves the study of sea water attention of electromagnetic (EM) signal propagation and the statistical interactions of the sea wave and LoS propagation.  


\subsection{Radio Wave Attenuation by Sea Water}
The sea water attention of radio signals is  investigated by examining the penetration capability of the radio wave into sea water. 
In this study, we adopt the ITU model \cite{ITU:Surface} for the electrical characteristics of the Earth’s surface, including pure water, sea water, ice, soil and vegetation cover, etc. The penetration depth, $\delta$, is defined as the depth at which the EM radiation amplitude falls to $1/e$ of the original value at the surface, in a homogeneous medium of complex relative permittivity $\varepsilon_{r}=\varepsilon^{'}_r-j\varepsilon^{''}_r$ 
and is given by~\cite{ITU:Surface} 
\begin{equation}\label{eq:Permittivity}
\delta=\frac{\lambda}{2\pi} \sqrt{\frac{2}{\sqrt{(\varepsilon^{'}_r)^2+(\varepsilon^{''}_r)^2}-\varepsilon^{'}_r}},
\end{equation}
where $\lambda$ is the radio wavelength in meters. Furthermore, considering the pure water model from ITU, the complex relative permittivity, $\varepsilon_\text{pw} = \varepsilon^{'}_\text{pw} - j\varepsilon^{''}_\text{pw}$, is a function depending on frequency $f_\text{GHz}$, and temperature $T$~($^{\circ}$C), as given by   
\begin{equation}\label{eq:Permittivity3}
\varepsilon^{'}_\text{pw}=\frac{ {\varepsilon_\text{s}-\varepsilon_\text{1}}} {1+(f_\text{GHz}/f_\text{1})^2}+\frac{ {\varepsilon_{\text{1}}-\varepsilon_{\infty}}} {1+(f_\text{GHz}/f_\text{1})^2}+\varepsilon_\infty,
\end{equation}
\begin{equation}\label{eq:Permittivity4}
\varepsilon^{''}_\text{pw}=\frac{ (f_\text{GHz}/f_\text{1})(\varepsilon_\text{s}-\varepsilon_\text{1})}{1+(f_\text{GHz}/f_\text{1})^2}+\frac{ (f_\text{GHz}/f_\text{2})(\varepsilon_\text{1}-\varepsilon_\infty)}{1+(f_\text{GHz}/f_\text{2})^2}, 
\end{equation}
where 
\begin{equation}\label{eq:Permittivity5}
\varepsilon_\text{s}=77.6+103.3\Theta,
\end{equation}
\begin{equation}\label{eq:Permittivity6}
\varepsilon_\text{1}=0.0671\varepsilon_\text{s},
\end{equation}
\begin{equation}\label{eq:Permittivity7}
\varepsilon_\infty=3.52-7.52\Theta,
\end{equation}
\begin{equation}\label{eq:Permittivity7}
\Theta=\frac{300}{T+273.15}-1,
\end{equation}
and $f_\text{1}$ and $f_\text{2}$ are the Debye relaxation frequencies expressed as
\begin{equation}\label{eq:Debye1}
f_\text{1}=20.2-146.4\Theta+316\Theta^2(\text{GHz}),
\end{equation}
\begin{equation}\label{eq:Debye2}
f_\text{2}=39.8f_\text{1}(\text{GHz}).
\end{equation}

\begin{figure}[t]
\begin{center}
\includegraphics[width=3.8in]{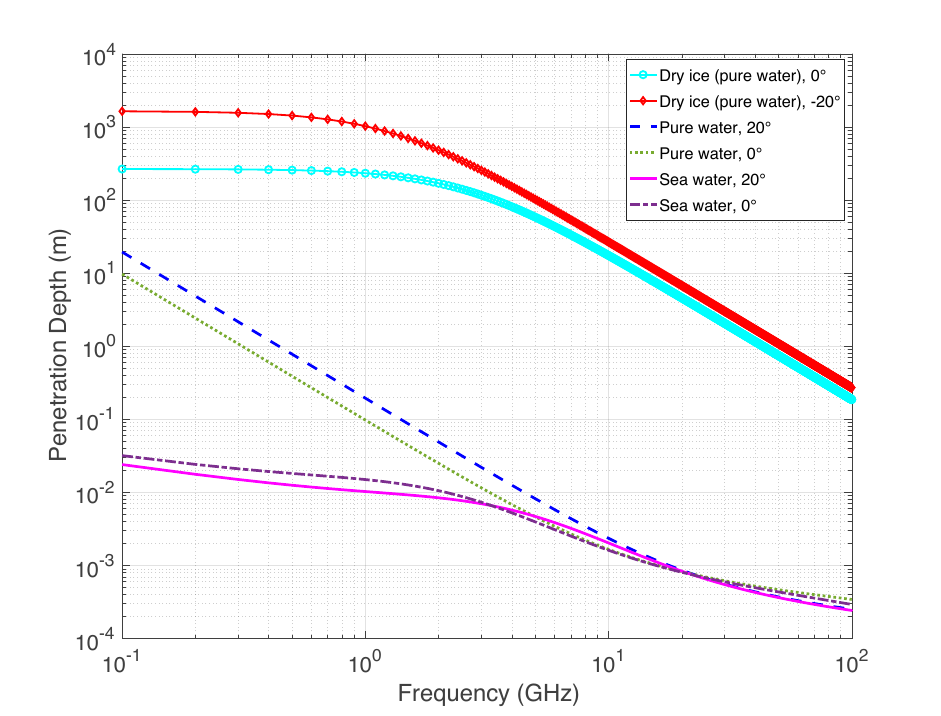}
\caption{Penetration depths of various surface types for radio wave propagation over 100 MHz to 100 GHz.}\label{fig:penetration}
\end{center}
\end{figure}

Furthermore, the sea (saline) water complex relative permittivity is a function of frequency $f_\text{GHz}$, temperature $T$~($^{\circ}$C), and salinity $S$ (g/kg or ppt), and can be derived as \cite{ITU:Surface}

\begin{equation}\label{eq:Permittivity8}
\varepsilon_{\text{sw}} = \varepsilon^{'}_{\text{sw}}-j\varepsilon^{''}_{\text{sw}},
\end{equation}
\begin{equation}\label{eq:Permittivity9}
\varepsilon^{'}_{\text{sw}}=\frac{ {\varepsilon_{\text{ss}}-\varepsilon_{\text{1s}}}} {1+(f_{\text{GHz}}/f_{\text{1s}})^{2}}+\frac{ {\varepsilon_{\text{1s}}-\varepsilon_{{\infty}_{\text{s}}}}} {1+(f_{\text{GHz}}/f_{\text{2s})^2}}+\varepsilon_{\infty_{\text{s}}},
\end{equation}

\begin{equation}\label{eq:Permittivity10}
\varepsilon^{''}_{\text{sw}}=\frac{ (f_{\text{GHz}}/f_{\text{1s}})(\varepsilon_{\text{ss}}-\varepsilon_{\text{1s}})}{1+(f_{\text{GHz}}/f_{\text{1s}})^{2}}+\frac{ (f_{\text{GHz}}/f_{\text{2s}})(\varepsilon_{\text{1s}}-\varepsilon_{\infty_{\text{s}}})}{1+(f_{\text{GHz}}/f_{\text{2s}})^2}+\frac{18\sigma_{\text{sw}}}{f_{\text{GHz}}}, 
\end{equation}
where
\begin{equation}\label{eq:Permittivity11}
\begin{split}
\varepsilon_{\text{ss}}=\varepsilon_{\text{s}}\exp(-3.56417\times10^{-3}S+4.74868\times10^{-3}S+\\
1.15574\times10^{-5}TS)
\end{split}
\end{equation}


\begin{equation}\label{eq:Permittivity13}
\begin{split}
f_\text{1s}=f_\text{1}(1+S(2.39357\times10^{-3}-3.13530\times10^{-5}T+\\
2.52477\times10^{-7}T^{2}))~(\text{GHz}),
\end{split}
\end{equation}

\begin{equation}\label{eq:Permittivity14}
\begin{split}
\varepsilon_\text{1s}=\varepsilon_\text{1}\exp(-6.28908\times10^{-3}S+1.76032\times10^{-4}S-\\
9.22144\times10^{-5}TS),
\end{split}
\end{equation}

\begin{equation}\label{eq:Permittivity15}
\begin{split}
f_\text{2s}=f_\text{2}(1+S(-1.99723\times10^{-2}+1.81176\times10^{-4}T))\\
(\text{GHz}),
\end{split}
\end{equation}
\begin{equation}\label{eq:Permittivity16}
\begin{split}
\varepsilon_{\infty_\text{s}}=\varepsilon_\infty(1+S(-2.04265\times10^{-3}+1.57883\times10^{-4}T)).
\end{split}
\end{equation}
The values of $\varepsilon_\text{s}$, $\varepsilon_\text{1}$, $\varepsilon_\infty$, $f_\text{1}$ and $f_\text{2}$ are obtained from~\eqref{eq:Permittivity5}-\eqref{eq:Debye2}. Furthermore, $\sigma_\text{sw}$ is given by
\begin{equation}\label{eq:Permittivity17}
\begin{split}
\sigma_\text{sw}=\sigma_\text{35}R_\text{15}R_\text{T15} (\text{S/m}),
\end{split}
\end{equation}
\begin{equation}\label{eq:Permittivity18}
\begin{split}
\sigma_\text{35}=2.903602 + 8.607 \times 10^{-2}T + 4.738817 \times 10^{-4}T^{2} \\
- 2.991 \times 10^{-6}T^{3} + 4.3047 \times 10^{-9}T^{4},  
\end{split}
\end{equation}
\begin{equation}\label{eq:Permittivity19}
\begin{split}
R_\text{15}=S (\frac{37.5109 + 5.45216S + 1.4409 \times 10^{-2}S^{2}}{1004.75 + 182.283 S + S^{2}}),
\end{split}
\end{equation}
\begin{equation}\label{eq:Permittivity20}
\begin{split}
R_\text{T15}=1 + \frac{\alpha_\text{0}(T-15)}{\alpha_\text{1}+T},
\end{split}
\end{equation}
\begin{equation}\label{eq:Permittivity21}
\begin{split}
\alpha_\text{0}=\frac{6.9431 + 3.2841S - 9.9486 \times 10^{-2}S^{2}}{84.850 + 69.024 S + S^{2}},
\end{split}
\end{equation}
\begin{equation}\label{eq:Permittivity22}
\begin{split}
\alpha_\text{1}=49.843 - 0.2276 S + 0.198 \times 10^{-2}S^{2}.
\end{split}
\end{equation}
When $S=0$ (for pure water), \eqref{eq:Permittivity9}-\eqref{eq:Permittivity10} are degenerated into \eqref{eq:Permittivity3}-\eqref{eq:Permittivity4}. On the other hand, considering the dry ice composed of pure water ($\leq 0~^{\circ}$C), the complex relative permittivity, $\varepsilon_{\text{ice}}=\varepsilon^{'}_{\text{ice}} - j \varepsilon^{''}_\text{ice}$, 
has the real part $\varepsilon^{'}_\text{ice}$ written as
\begin{equation}\label{eq:Permittivity24}
\begin{split}
\varepsilon^{'}_\text{ice}=3.1884 + 0.00091T, 
\end{split}
\end{equation},
which is a function of temperature $T$, and independent of frequency $f_\text{GHz}$.
The imaginary part $\varepsilon^{''}_\text{ice}$ is a function of $T$ and frequency $f_\text{GHz}$ given by
\begin{equation}\label{eq:Permittivity25}
\begin{split}
\varepsilon^{''}_\text{ice}=A/f_\text{GHz} + Bf_\text{GHz}, 
\end{split}
\end{equation}
where we further have
\begin{equation}\label{eq:Permittivity26}
\begin{split}
A=(0.00504 + 0.0062\Theta)\text{exp}(-22.1\Theta),
\end{split}
\end{equation}
\begin{equation}\label{eq:Permittivity27}
\begin{split}
B=\frac{0.0207}{T+273.15}\frac{\exp(-\uptau)}{(\exp(-\uptau)-1)^{2}} + 1.16 \times 10^{-11}f^{2}_\text{GHz} +\\
\exp(-9.963 + 0.0372 T),
\end{split}
\end{equation}
\begin{equation}\label{eq:Permittivity28}
\begin{split}
\uptau=\frac{335}{T+273.15},
\end{split}
\end{equation}
\begin{equation}\label{eq:Permittivity29}
\begin{split}
\uptau=\frac{300}{T+273.15}-1.
\end{split}
\end{equation}
Finally, by using above equations, the penetration depths of radio wave propagation from 100 MHz to 100 GHz for pure water, sea water (35 ppt), and dry ice at different temperatures, are plotted in Fig. \ref{fig:penetration}. As observed, sea water at 20$^{\circ}$C results in the smallest penetration depth (meaning the most significant attenuation) for frequencies below 3 GHz, while dry ice at -20$^{\circ}$C demonstrates the least attenuation at frequencies of interest. At 1 GHz, the penetration depth of sea water and dry ice is 0.01 m and 1000 m, respectively. 

\begin{figure}[t]
\begin{center}
\includegraphics[width=3.3in]{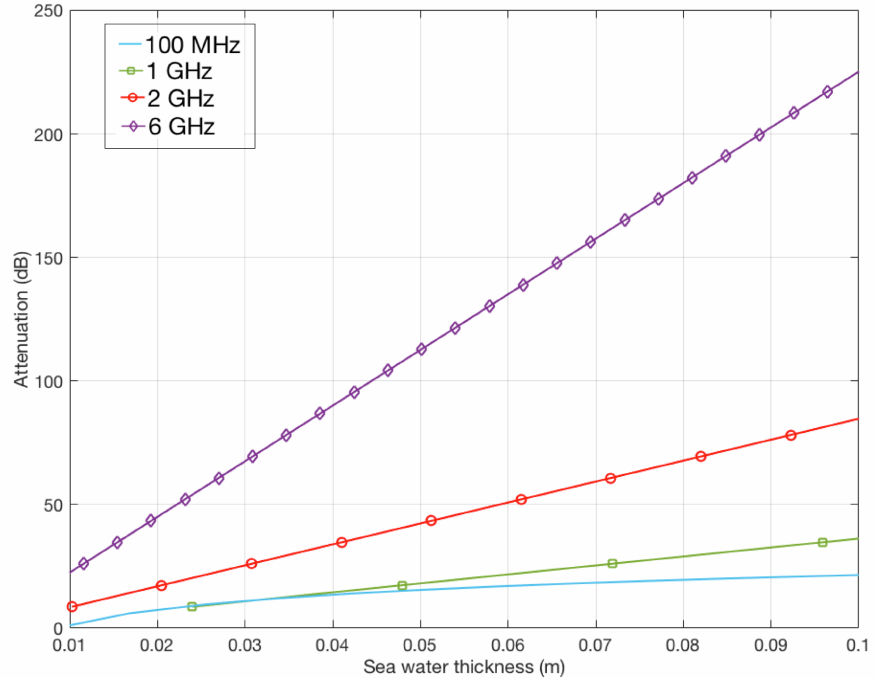}
\caption{Radio wave attenuation by sea water as a function of thickness.}\label{fig:attenuation}
\end{center}
\end{figure}

Based on the simulated penetration depths, the radio wave attenuation as a function of the sea water thickness for carrier frequencies of interest such as 100 MHz, 1 GHz, 2 GHz and 6 GHz, is further simulated and illustrated in Fig. \ref{fig:attenuation}. The sea water of around 0.1 m thick can result in more than 34 dB and 78 dB attenuation at 1 GHz and 2 GHz, respectively. In other words, such large attenuation can lead to communication outage or deteriorate the communication links when ocean waves block the communication paths.

\begin{figure*}[t]
\begin{center}
\includegraphics[width=6.8in]{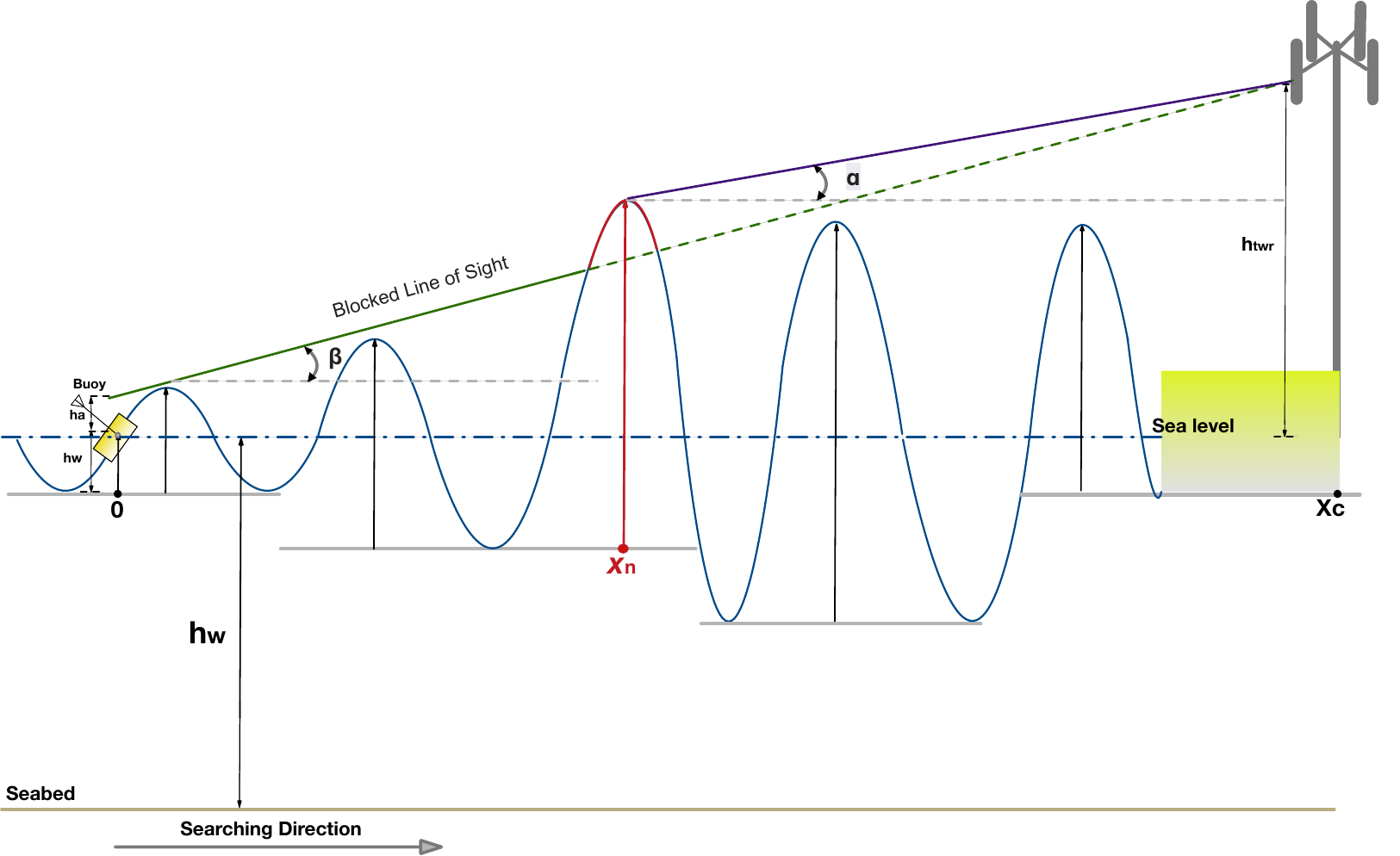}
\caption{Illustration of buoy-to-land communication in the presence of sea waves.}\label{fig:OTS}
\end{center}
\end{figure*}

 Fig.~\ref{fig:OTS} illustrates a buoy-to-land communication over the sea, when ocean waves of sizable thickness (related to the wavelength of the ocean wave) get in the way between the buoy antenna and the cell tower, and the LoS path is severely blocked. Furthermore, as the sea surface elevation dynamically changes and alternatively enables and disables the LoS communication, a detailed study involving ocean wave modeling is necessary.

\subsection{Wave Modeling in Coastal and Oceanic Waters}
The actual mechanism of fluid dynamics of the ocean wave and the surface elevation variation in a highly dynamic oceanic environment is very complicated. 
Therefore, it is important to investigate how precisely the surface elevation of the sea water varies over time through statistical modeling. \textcolor{black}{With regard to the ocean wave generation, there is a high degree of correlation between wind speed and wave height. In terms of the wave status, there are mainly four types known as, starting seas, developing seas, fully developed seas, and decaying seas. For the wave spectra analysis, Pierson-Moskowitz (P-M) spectrum was introduced in 1964 by offshore industry as one parameter spectrum based for fully developed seas in the North Atlantic generated by local winds \cite{Hol:Waves}. However, P-M spectrum is limited for analysis since its need for fully developed seas is too restrictive \cite{Techet:Waves}. The two parameter spectrum,} Bretschneider spectrum, also known as ISSC spectrum (represented by significant wave height and mean period) \cite{American:Waves} is the spectrum model recommended for open-ocean wave conditions given by \cite{Hol:Waves}, \cite{Techet:Waves} 
\begin{equation}\label{eq:ISSC}
\begin{split}
S_\text{$\eta$}(\omega) = \frac{5}{16}\frac{H^{2}_{s}\omega^{4}_{p}}{\omega^{5}} \exp{(-\frac{5}{4}(\frac{\omega_{p}}{\omega})^{4})}~\text{in}~\text{m}^{2}/(\text{rad/s}),
\end{split}
\end{equation}
where $H_{s}$ is the significant wave height (SWH) in meter, also known as $H_\text{1/3}$ which is defined traditionally as the mean wave height (trough to crest) of the highest third of the waves, and $\omega_{p}$ is the modal (peak) angular frequency in rad/s. Moreover, the peak period, $T_{p}$ = 2$\pi/\omega_{p}$.

\begin{figure}[t]
\begin{center}
\includegraphics[width=3.8in]{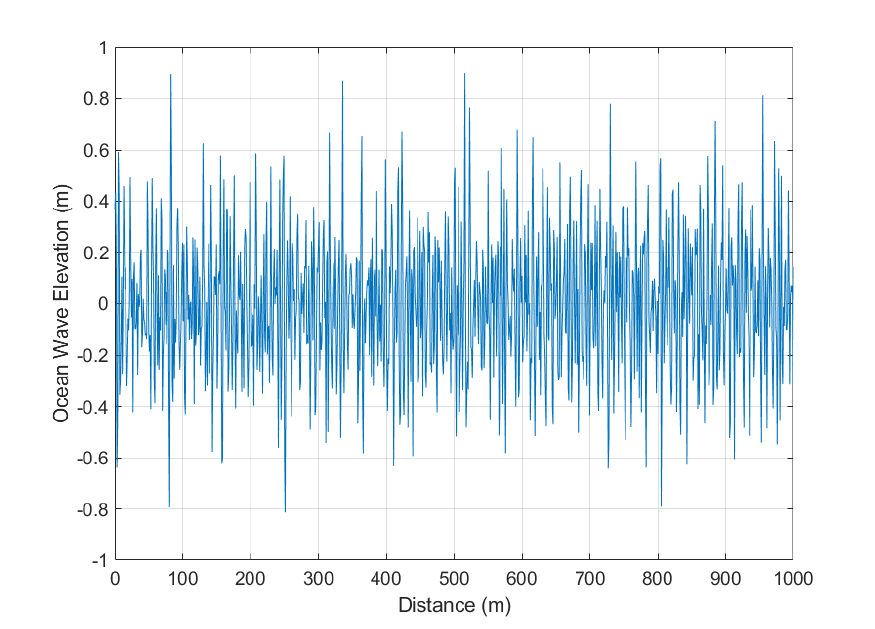}
\caption{Ocean wave elevation over distance for ($H_{s}$, $T_{p}$) at (1 m, 2 s).}\label{fig:ocean}
\end{center}
\end{figure}

\begin{figure}[t]
\begin{center}
\includegraphics[width=3.8in]{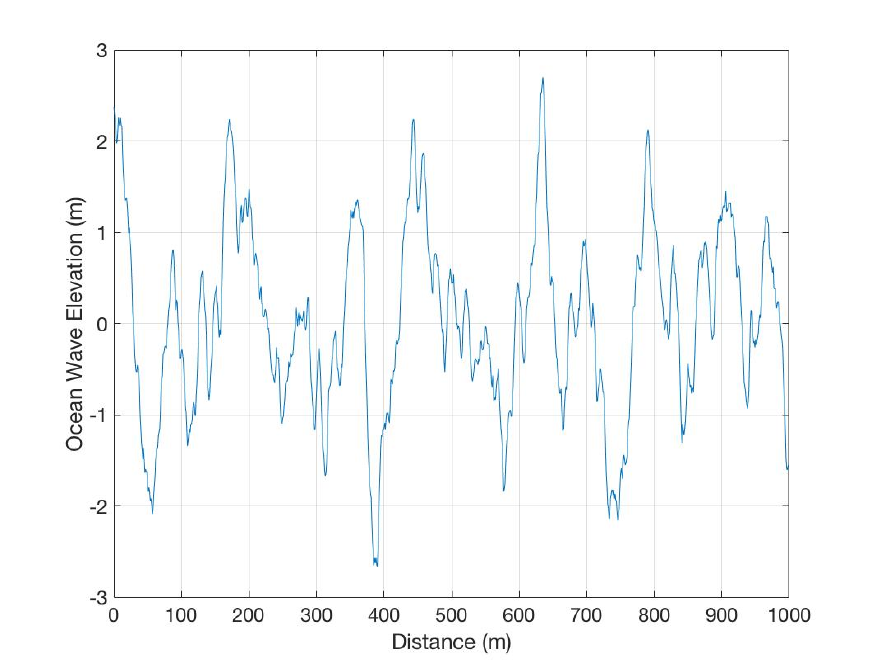}
\caption{Ocean wave elevation over distance for ($H_{s}$, $T_{p}$) at (4 m, 10 s).}\label{fig:ocean2}
\end{center}
\end{figure}

\begin{figure}[t]
\begin{center}
\includegraphics[width=3.8in]{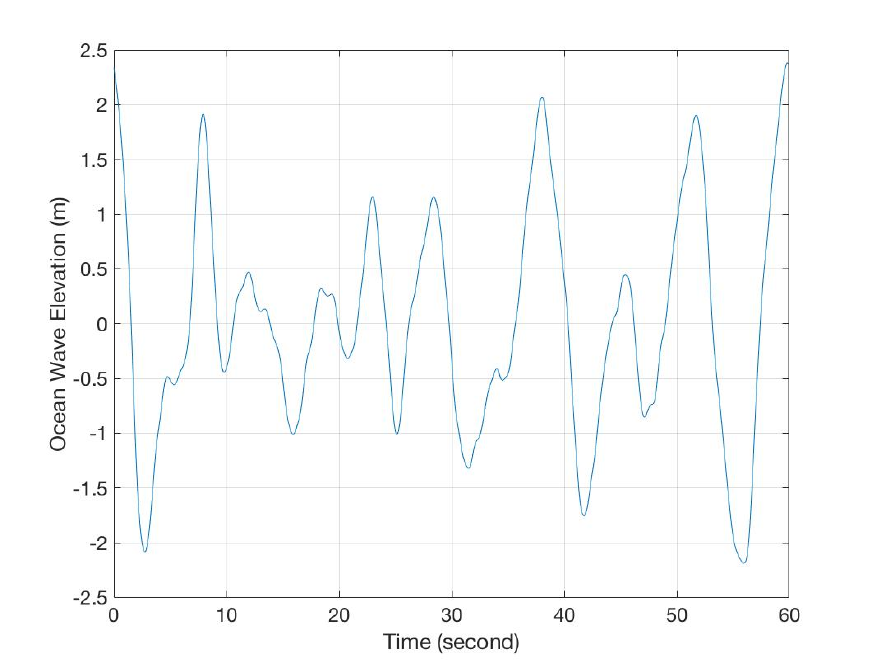}
\caption{Ocean wave elevation over 1-minute period for ($H_{s}$, $T_{p}$) at (4 m, 10 s).}\label{fig:ocean3}
\end{center}
\end{figure}

For a regular, monochromatic ocean wave of amplitude $A$, angular frequency $\omega$ and a phase constant $\varepsilon$, the moving ocean surface elevation $\eta(t)$ can be described as
\begin{equation}\label{eq:ISSC2}
\begin{split}
\eta(x, y, t) = \Re\big\{A\exp{(j*(-kx {\cos\theta}-ky {\sin\theta}+\omega t+ {\varepsilon}))}\big\},
\end{split}
\end{equation}

where $\theta$ is the direction \textcolor{black}{of wave propagation from the $x$ axis. For the case where $\varepsilon=0$, and no directionality is considered when the ocean propagation direction is normal to the shore~\cite{Beatty:Thesis},} \eqref{eq:ISSC2} is further reduced to
\begin{equation}\label{eq:ISSC3}
\begin{split}
\eta(x, t) = \Re\big\{A\exp{(j*(-kx+\omega t))}\big\}\\
= A\cos(\omega t-kx).
\end{split}
\end{equation}
If observing the water surface variation at the origin (also the location of the buoy) by making $x=0$, we have
\begin{equation}\label{eq:ISSC4}
\begin{split}
\eta(0, t) = A\cos(\omega t).
\end{split}
\end{equation}
Note that the actual ocean wave is composed of a large number, $N_f$, of frequency components. Therefore, the ocean wave is the summation of ocean waves of all frequency components given by~\cite{Hol:Waves} 
\begin{equation}\label{eq:ISSC5}
\begin{split}
\eta(x,t) = \sum_{i=1}^{N_f} a_{i}\cos(2\pi f_{i}t + k_{i}x + \alpha_{i}),
\end{split}
\end{equation}
where $a_{i}$, $k_{i}=\frac{2\pi}{\lambda_{i}}$ and $\alpha_{i}$ are the amplitude, wave number, and phase for the $i$-th frequency component $f_{i}$. In sufficiently deep water where water depth $h_{w}$ $>$ 0.3$\lambda$ \cite{Beatty:Thesis}, $k_{i}$ is given by $k_{i}=\omega_{i}^{2}/g$, where $g$ is the gravity constant. Moreover, each corresponding amplitude component $a_{i}$ follows Rayleigh distribution with probability density function (PDF) given by
\begin{equation}\label{eq:ISSC6}
\begin{split}
p_a(a_{i})=\frac{\pi}{2}\frac{a_{i}}{\mu^{2}_{i}}\exp{(-\frac{\pi a^{2}_{i}}{4\mu^{2}_{i}})}~\text{for}~ a_{i} \geq 0, 
\end{split}
\end{equation}
where $\mu_{i} = E\big\{a_{i}\big\}$ is the expected amplitude value and $\alpha_{i}$ is the phase uniformly distributed between 0 and 2$\pi$ with PDF
\begin{equation}\label{eq:ISSC7}
\begin{split}
p_{\alpha}(\alpha_{i})=\frac{1}{2\pi}~\text{for}~0 < \alpha_{i} \leq 2\pi. 
\end{split}
\end{equation}
Furthermore, the expected amplitude $\mu_{i}$ of each frequency component can be calculated as~\cite{Beatty:Marine}, \cite{Beatty:Thesis}
\begin{equation}\label{eq:ISSC8}
\begin{split}
\mu_{i}=\sqrt{2\cdot S_\eta(\omega_{i})\cdot \Delta\omega},
\end{split}
\end{equation}
where $S_\eta(\omega_{i})$ is the ocean wave spectra that can be obtained using \eqref{eq:ISSC}, and $\Delta\omega$ is the frequency bin width of the spectrum $S_\eta(\omega)$. Therefore, at the buoy's location, i.e., $x=0$, we have
\begin{equation}\label{eq:ISSC9}
\begin{split}
\eta(0, t) = \sum_{i=1}^{N_f} a_{i}\cos(\omega_{i} + \alpha_{i}).
\end{split}
\end{equation}
Moreover, at any other location, e.g., $x_{p}$, we further have
\begin{equation}\label{eq:ISSC10}
\begin{split}
\eta(x_{p}, t) = \sum_{i=1}^{N_f} a_{i}\cos(\omega_{i} + k_{i}x_{p} + \alpha_{i}).
\end{split}
\end{equation}
Consequently, once the combination of ($H_{s}$, $T_{p}$) is defined, by using \eqref{eq:ISSC6}--\eqref{eq:ISSC10}, we can simulate the time-varying ocean wave in any location between the buoy and cellular tower in the time domain, with adjustable time interval as tunable resolution. In particular, according to the historical data, field test and ocean engineering experience, the generated $a_{i}$  and $\alpha_{i}$ from one random realization can be used for simulating, for example, 1-minute duration ocean wave. Then new $a_{i}$ and $\alpha_{i}$ are generated from a new random realization every 1 minute to emulate the real coastal waves. 
As examples, the ocean waves over the 1000-meter distance at some time instants are generated and plotted for two combinations of ($H_{s}$, $T_{p}$), respectively. In Fig.~\ref{fig:ocean} where $H_{s}$ is 1 meter and $T_{p}$ is 2 second, at one time instant along its propagation direction, the ocean wave elevation (OWE) varies between $-0.8$ to $+1$ m; while in Fig.~\ref{fig:ocean2}, the OWE varies between $-3.6$ to $+3.2$ m for $(H_{s}, T_{p})$ at (4 m, 10 s). Furthermore, if we fix the observation point to the buoy’s location, the OWE over 1-minute period is plotted in Fig.~\ref{fig:ocean3}. 

To summarize, the movement and mobility in the ocean environment can significantly affect the communication conditions. 
Usually, when an ocean wave blocker appears between the buoy and the tower as shown in Fig.~\ref{fig:OTS}, the LoS communication is unavailable at cellular frequencies of interest, due to the very strong attenuation caused by penetration loss. Nevertheless, the specular path or other non-line-of-sight (NLoS) diffuse paths in this case may still exist. But it is difficult to characterize such NLoS communications and conduct the quantitative analysis, due to the dynamic ocean environment and complicated ocean wave reflection and scattering scenarios. Moreover, since such multi-path components (MPCs) may be very random and unreliable, the received power at the tower base station is unpredictable.
Therefore, for the feasibility of analysis and practical deployment reliability, when the ocean wave blocking happens, there is neither actual LoS nor effective NLoS communications between the buoy and the tower. 

\begin{algorithm}[t!]
{\footnotesize\caption{\footnotesize Ocean Wave Blocker Searching and LoS Communication Probability Characterization}
\label{alg:proposed_algorithm}
    \hspace*{\algorithmicindent} \textbf{Input: }($H_{s}$, $T_{p}$), $H_{twr}$, $h_{a}$, $d$ and $N_{rp}$  \\
    \hspace*{\algorithmicindent} \textbf{Output:} $P_{LoS}$, statistical distribution of continuous LoS communications with $\mu_{CLoS}$ and $\sigma_{CLoS}$ 
	\begin{algorithmic}[1]
		\STATE Generate wave number $k_{i}$ for each frequency component $\omega_{i}$ by setting $k_{i}$:=$\omega^{2}_{i}$/g; set $N_{a}$:=0; set $N_{f}$:= number of frequency components; 
		\WHILE {$N_{a}<N_{rp}$} 
		\FOR {$t = 0: \Delta t: T$ ($T$ is the period of one realization for the ocean wave, 60 seconds, $\Delta t$ is set to 0.1 second) }
		\STATE Set $\mu_{i}:= \sqrt{2\cdot S_{\eta}(\omega_{i})\cdot \Delta \omega }$ \newline
		Set $a_{i}:= \text{raylrnd} (\mu_{i}$) \newline
		Set $\alpha_{i}:= \text{rand} (1, N_{f})\times 2 \times \pi$ \newline
		$\eta(0, t) = \sum_{i=1}^{N_f} a_{i}\cos(\omega_{i} + k_{i}x + \alpha_{i})$ \newline
		$\eta(x_{n}, t) = \sum_{i=1}^{N_f} a_{i} \cos(\omega_{i} + k_{i} x_{n} + \alpha_{i})$ 
		\FOR{$x_n = 1: d$ (search the blocker in current time instance)}
        \IF{$\frac{h_{twr}-(\eta(0, t)+h_{a})}{d}>\frac{h_{twr}-\eta(x_{n}, t)}{d-x_n}$} 
        \STATE record the distance $d_{blk}=x_n$ and height of the blocker \newline
        break
        \ENDIF
		\ENDFOR
		\ENDFOR
\textcolor{black}{		\STATE $N_{a}$:=$N_{a}$ + 1 \newline
     	regenerate a realization set of $a_{i}$ and $\alpha_{i}$ randomly }
		\ENDWHILE
\end{algorithmic}}
\end{algorithm}

\textcolor{black}{On the other hand, it is worth mentioning that the ducting effect caused by the refractivity change (due to the change of atmospheric pressure, temperature, etc.) at different heights of the atmosphere \cite{Jin:Maritime} includes the evaporation duct that can be utilized for beyond LoS (B-LoS) maritime communications. Moreover, the evaporation duct over the sea surface has an appearance height around 10 m -- 20 m (at most 40 m) and can trap the EM waves between the ducting layer and the sea surface, which can enable the signals travel over the horizon without spreading isotropically \cite{Dinc:Surface2}. In addition, the evaporation duct appearance is highly related to the climate, season and other special geographical conditions. For example, in the equatorial and tropical areas, the appearance probability is as high as 90\%. In a further communication experiment conducted in the Great Barrier Reef of Australia, the atmospheric ducts can connect a 78-km link at 10.5 GHz with 10 Mb/s speed and 80 percent of the time \cite{Woods:Duct}. However, the evaporation duct enabled B-LoS channels demonstrate several key features, such as: 1) X-band (8.0 -- 12.0 GHz) is more favorable than other bands to enhance the communication link; 2) Only transmission within a certain angle range can be trapped in the duct layer.
Considering that in our research, sub-6 GHz frequency bands (e.g. LTE Band 1/4 and 5G low/mid bands) are used and the buoys are deployed along the coastal line of the Pacific (above 48° north latitude), the evaporation duct appearance probability is rather tiny and therefore it will not be taken into account in this paper. }

Consequently, the probabilistic LoS channel model plays a crucial role in the scope of this research, and it can be obtained based on the ocean wave modeling, which indicates that the probability of LoS communications, $P_{LoS}$, is a function of ($H_{s}$, $T_{p}$). In order to facilitate energy-efficient maritime IoT communications, the first step is to obtain the LoS communication statistics under various ocean environments.

\subsection{Probability of LoS and Continuous LoS Duration}
Using ocean wave modeling, we can generate many wave realizations through simulations. A geometric condition can be formulated and examined for each realization to determine if this particular waveform blocks the LoS link between the buoy antenna and the cell tower antenna. As shown in Fig.~\ref{fig:OTS}, when the LoS communication is disabled by an ocean wave blocker appearing at $x_{n}$, there is $\beta>\alpha$, where $\beta$ is the angle between the LoS and the horizontal sea level and $\alpha$ is the angle of the line connecting the sea surface at $x=x_n$ and the cell tower antenna, with respect to the sea level. This relationship leads to the following wave blocking criteria
\begin{equation}\label{eq:BSA}
\begin{split}
\frac{h_{twr}-(\eta(0,t)+h_{a})}{d}>\frac{h_{twr}-\eta(x_{n}, t)}{d-x_{n}},
\end{split}
\end{equation}
where $h_{a}$ is the effective antenna height of the buoy and it is vertically measured from the sea level, $h_{twr}$ is the height of the cellular tower, $d$ is the distance between buoy and the cellular tower, $x_{n}$ is the horizontal location of the ocean wave blocking point, and $\eta(0, t)$ and $\eta(x_{n},t)$ are the ocean surface elevation at the location of the buoy and the ocean wave blocker, respectively. As illustrated in Fig.~\ref{fig:OTS}, $d=x_{c}$, $h_{a}$ = $L_{a}$sin$\beta_{a}$, $L_{a}$ is the physical dimension of the antenna, and $\beta_{a}$ is the tilt angle formed between the antenna and sea level. Considering $L_{a}$cos$\beta_{a} \ll x_{c}$, we can assume that buoy’s antenna and buoy are located at the same horizontal origin.

An iterative algorithm of locating the nearest ocean wave blocker(s) and analyzing the average LoS probability and the stochastic continuous LoS segments is summarized in Algorithm 1, where $N_{rp}$ denotes the number ($>$1,000) of executed random realizations each of which simulates the dynamic ocean surface between the buoy and the tower, within a $T$-length  
time window. The temporal resolution (step), $\Delta t$, should be set to 
several times smaller than the reciprocal of the largest frequency component $f_{N}$ (1.5 Hz), ensuring that the high-accuracy simulation results can be maintained without requiring challenging computing capability. Within one random realization, for every time instant, the blocker search is performed over the distance between the buoy and the tower, and the location and height of the ocean wave blocker are also recorded. For the $T$-length time window, a total of $N=T/\Delta t$ wave snapshots between the buoy and the tower are generated and searched. After that, $a_{i}$, and $\alpha_{i}$ are regenerated to get another realization, and a total of $N_{rp}$ set of realizations are run to collect statistically consistent results.        

Dynamic wave blocking results in interrupted LoS connections. To quantify, denote the continuous LoS (CLoS) segments within a $T$ second window having time intervals $t_{1}$, $t_{2}$, $\ldots$, $t_{M}$, and the continuous blocked LoS (BLOS) segments with the time duration $\bar{t}_{1}$, $\bar{t}_{2}$, {\ldots}, $\bar{t}_{\bar{M}}$.
The total LoS time duration is defined as $T_{LoS}=\sum_{m=1}^{M} t_m$. The probability of LoS communication, $P_{LoS}$ is calculated by
\begin{equation}\label{eq:PLOS}
\begin{split}
P_{LoS}=\frac{T_{LoS}}{T}=\frac{\sum_{m=1}^{M} t_m}{T},
\end{split}
\end{equation}
and $P_{BLoS}=1-P_{LoS}$. In simulation $P_{LoS}$ is averaged over all $N_{rp}$ realizations. 

To characterize the statistical behavior of CLoS duration, $t_{CLoS}$, we define $N$ interval bins for the $T$ duration, where the $k$-th bin spans  $(t^k-\Delta t, t^k]$ and $t^k=k\Delta t$, $k=1,\ldots,N$. The histogram count in the $k$-th bin for $t_{CLoS}$, $h_k$, is given by
\begin{equation}
h_k=\sum_{m=1}^{M} I(t_m \in (t^k-\Delta t, t^k])
\end{equation}
where $I(\cdot)$ is the indicator function that returns value 1 if the inside set condition is met and value 0 otherwise. The histogram of BLoS can be obtained in a similar manner.

We also define a parameter $P_T^k$, the percentage of time occupied by CLoS segments of length $t^k$ over the total LoS duration, written as
\begin{equation}
P_T^k=\frac{h_k t^k}{\sum_{i=1}^{N}h_i t^i}    
\end{equation}
where $\sum_{i=1}^{N}h_i t^i$ is the total LoS time duration $T_{LoS}$. Next we use the conditional complementary cumulative distribution function (CCDF) of CLoS duration to characterize the conditional probability that the CLoS duration is larger than a threshold. This helps in practice to estimate the success rate of data packet transmission with a certain packet length. That is,
\begin{equation}\label{eq:CCDF}
\begin{split}
F(t^k|LoS)=P(t_{CLoS}\geq t^k|LoS)= \frac{\sum_{i=k}^N h_i}{\sum_{i=1}^N h_i}.
\end{split}
\end{equation}

Finally, suppose the minimum time interval required for transmission of a data packet in a communication system is $T_h$, the \textcolor{black}{LoS outage probability} that such transmission fails due to sea wave dynamics can be given by
\begin{equation}\label{eq:Pout}
\begin{split}
P_{out}(T_h)=P_{BLoS}+P_{LoS}P(t_{CLoS}<T_h|LoS)\\
=P_{BLoS}+P_{LoS}(1-F(T_h|LoS)).
\end{split}
\end{equation}
 
\subsection{Numerical Analysis}
\begin{figure}[t]
\begin{center}
\includegraphics[width=3.2in]{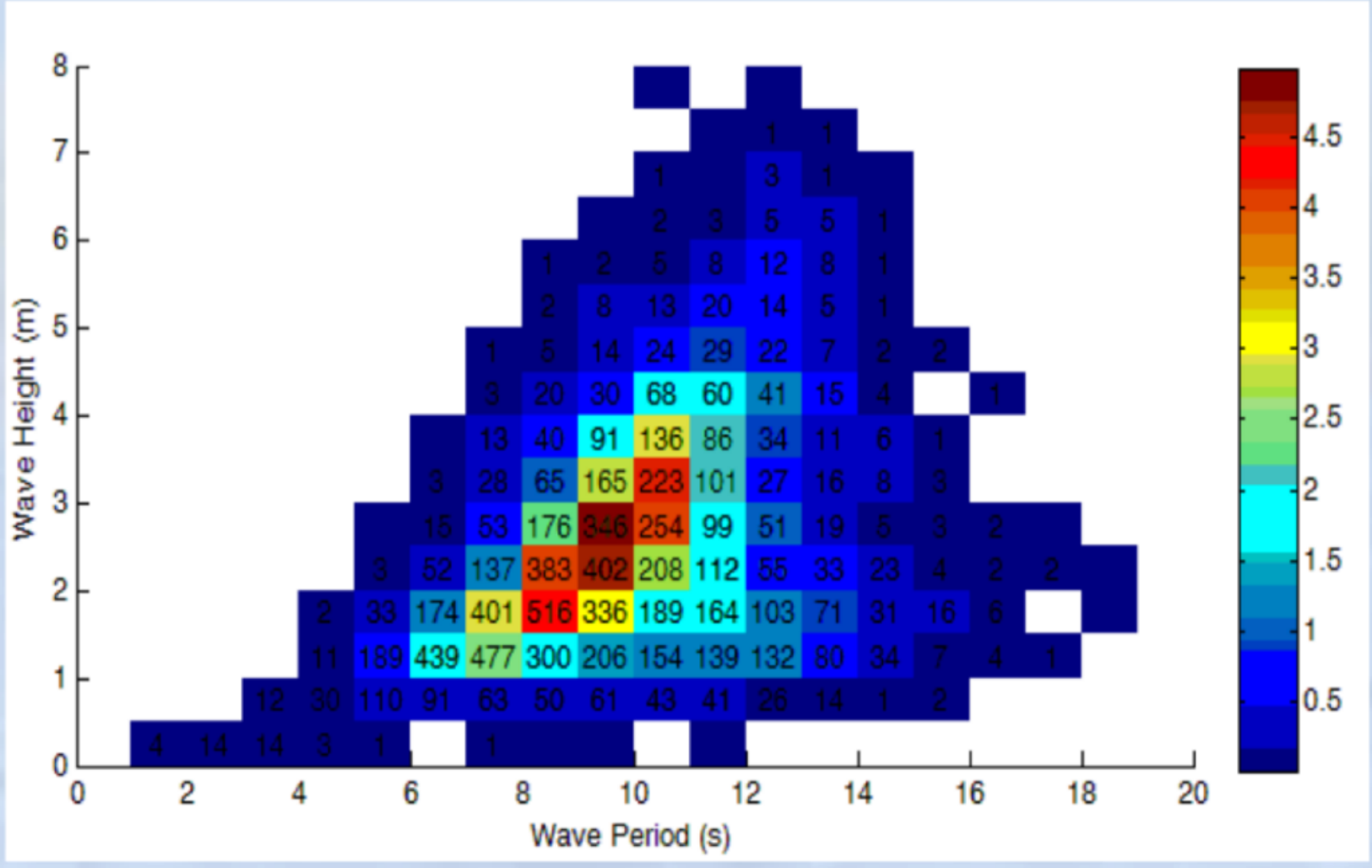}
\caption{Scatter diagram of occurrences in a full year of Tofino, BC~\cite{Beatty:Thesis}.}\label{Fig:Scatter}
\end{center}
\end{figure}

Among the input parameters, $H_{twr}$, $h_{a}$, and $d$ can be considered as design parameters, while ($H_{s}$, $T_{p}$) describe environment conditions. In particular, the effective antenna height, $h_{a}$, is directly relevant to the maritime IoT system design and performance. The distance between the buoy and the tower, $d$, is also interesting to investigate since it directly influences the  deployment strategy. The choice of $(H_s, T_p)$ combinations is guided by the historical measurement data visualized in Fig.~\ref{Fig:Scatter}, a scatter diagram of wave occurrences in a full year of Tofino,  BC~\cite{Beatty:Thesis}. When the amplitude of a wave reaches a critical level at which large amounts of wave energy is transformed into turbulent kinetic energy, the wave breaking phenomenon happens. In this scenario, simple physical models that describe wave dynamics with linear behaviour assumed become invalid. Therefore, all cases with $H_{s} > 0.8T_{p}$ are rejected and not simulated, and hence $T_{p}$ starts with 2 s and 6 s for $H_{s}$~at 0.5 m and 4 m, respectively.

\begin{figure}[t]
\begin{center}
\includegraphics[width=3.6in]{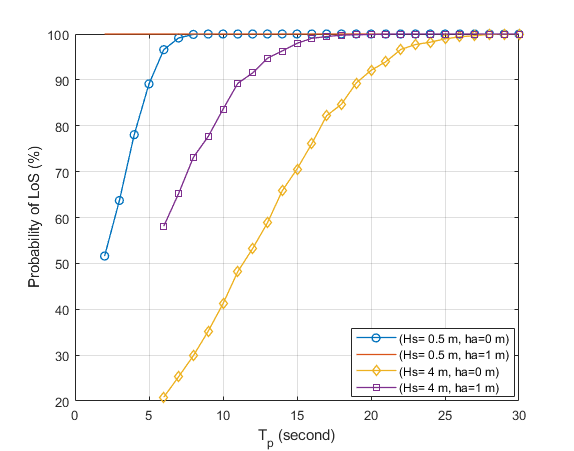}
\caption{Probability of LoS communication over $T_{p}$ for $H_{s}$ at 0.5 m and 4 m, when the effective antenna height is 0 and 1 m, respectively.}\label{fig:ocean4}
\end{center}
\end{figure}



\begin{figure}[t]
\begin{center}
\includegraphics[width=3.6in]{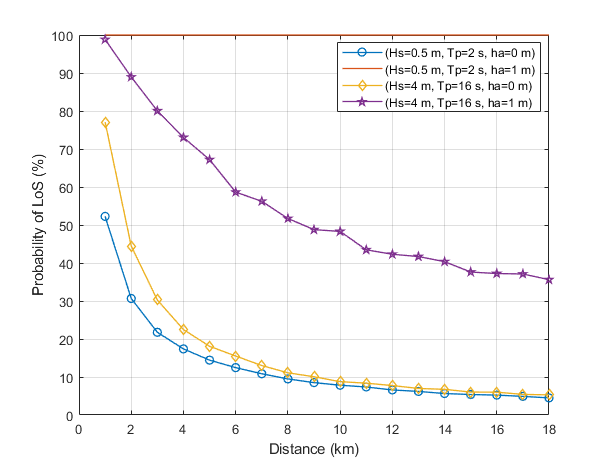}
\caption{Probability of LoS communication over distance for ($H_{s}$, $T_{p}$) at (0.5 m, 2 s) and (4 m, 16 s) when the effective antenna height $h_{a}$ is 0 and 1 m, respectively.}\label{fig:LoS_D}
\end{center}
\end{figure}

In our simulation, $T$ is set to 60 seconds and $\Delta t$ is 0.1 second. For each combination of the test scenario, at least 1000 realizations have been executed to obtain statistically meaningful results of interest. As illustrated in Fig.~\ref{fig:ocean4}, the probability of LoS communication over $T_{p}$ for $H_{s}$ at 0.5 m and 4 m are presented, respectively.  
It is observed that, $P_{LoS}$ increases with $T_{p}$ for all scenarios. Moreover, increasing the effective antenna height can significantly boost up $P_{LoS}$. 
For example, for the ($H_{s}$, $T_{p}$) combination of (0.5 m, 2 s), $P_{LoS}$~is almost doubled from 50\% to 100\% when increasing $h_{a}$ from 0 to 1 m. Moreover, $P_{LoS}$ is increased from 22\% to 58\% at the start value of $T_{p}$ when $h_{a}$ changes from 0 to 1 m. For other ($H_{s}$, $T_{p}$) combinations, significant improvement can also be observed when the antenna height is increased. Fig.~\ref{fig:LoS_D} shows that the LoS probability decreases with distance while a higher antenna can substantially improve $P_{LoS}$, especially at large distances. 

\begin{figure}[t]
\begin{center}
\includegraphics[width=3.4in]{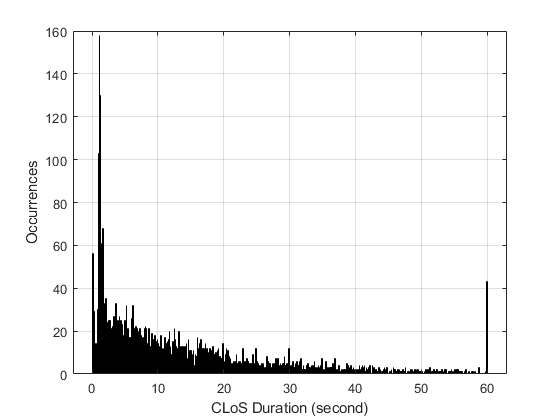}
\caption{Occurrence of the continuous LoS duration for ($H_{s}$, $T_{p}$, $h_{a}$) at (0.12 m, 2 s, 0 m).}\label{fig:CLoS}
\end{center}
\end{figure}

\begin{figure}[t]
\begin{center}
\includegraphics[width=3.4in]{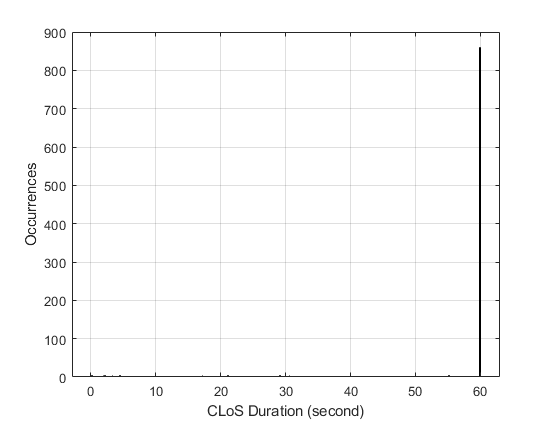}
\caption{Occurrence of the continuous LoS duration for ($H_{s}$, $T_{p}$, $h_{a}$) at (0.67 m, 2.8 s, 0.8 m).}\label{fig:CLoS2}
\end{center}
\end{figure}


In addition to the probability of LoS, characterizing the continuous LoS duration provides many insights for energy-efficient system and protocol design, such as data packing and segmentation, transmission scheduling, etc., to increase the burst data success rate and reduce retransmission. Retransmission is a major source of power drain.

\textcolor{black}{Figs.~\ref{fig:CLoS} and \ref{fig:CLoS2} show the occurrence of the continuous LoS duration in different $H_{s}$ and $T_{p}$ combinations. $H_{s}=0.12$ m and $T_{p}=2$ s are the mean values of the 41-day-20-hour measurement dataset from the MarineLabs CoastScout near Tofino, BC, Canada in 2019. The effective antenna height $h_{a}=0$ m. Since the temporal resolution of CLoS is 0.1 second and $T=60$ s, there are 600 CLoS bins ranging from 0.1-s CLoS duration to 60-s CLoS duration. The upper limit is capped to the examined time window but can be extended by using a large $T$. In practice, the choice of $T$ can be determined by the desired continuous LoS duration that can complete data packet transmission. According to the numerical result based on 1000 random realizations, there are 158/130/103/68/61/56 CLoS segments lasting for 1.1/1.2/1.0/1.6/1.4/0.1 seconds, respectively. Moreover, there are 43 CLoS segments lasting for 60 seconds. The mean value of the CLoS time interval, $\mu_{CLoS}$, is 12.98 seconds while the standard deviation, $\sigma_{CLoS}$, is 13.07 seconds. The probability of LoS for this case, $P_{LoS}$, is 98.59\%. In Fig.~\ref{fig:CLoS2},  $H_{s}$ is set to 0.67 m, the maximum value from the same measurement dataset. For ($H_{s}$, $T_{p}$, $h_{a}$) at (0.67 m, 2.8 s, 0.8 m), there are 859 CLoS communication lasting for $60$ seconds. The mean value of the CLoS time window, $\mu_{CLoS}$, is 51.37 seconds while the standard deviation, $\sigma_{CLoS}$, is 17.09 seconds, and $P_{LoS}$ is 99.92\%. In this combination, $h_{a}$ is a bit higher than $H_{s}$. } 

\begin{figure}[t]
\begin{center}
\includegraphics[width=3.4in]{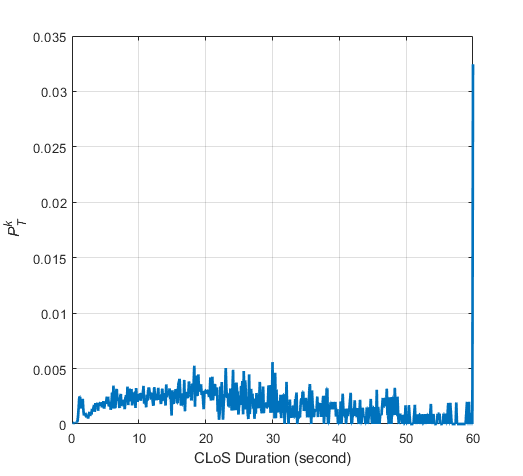}
\caption{$P_T^k$ for ($H_{s}$, $T_{p}$, $h_{a}$) at (0.12 m, 2 s, 0 m).}\label{fig:P_T_K}
\end{center}
\end{figure}

\begin{figure}[t]
\begin{center}
\includegraphics[width=3.4in]{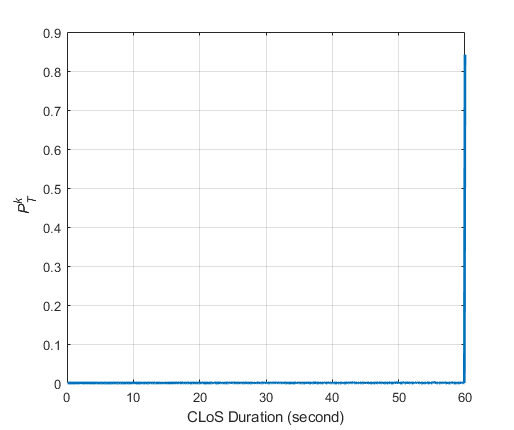}
\caption{$P_T^k$ for ($H_{s}$, $T_{p}$, $h_{a}$) at (0.67 m, 2.8 s, 0.8 m).}\label{fig:P_T_K2}
\end{center}
\end{figure}

\textcolor{black}{The percentage of time, $P_{T}^{k}$, occupied by CLOS equal to $t^k=k\Delta t$ over the total LoS time are plotted in Figs.~\ref{fig:P_T_K}~--~\ref{fig:P_T_K2}, for two test combinations at (0.12 m, 2 s, 0 m) and (0.67 m, 2.8 s, 0.8 m), respectively.}

Table~\ref{tab:marine} summarizes the CLoS results for more combinations. In addition to the mean and standard deviation of CLoS duration, $\gamma_{CLoS}$ represents the most probable CLoS duration which is the $t^{k}$ that corresponds to the highest $h_{k}$. In the scenario where the ($H_{s}$, $T_{p}$, $h_{a}$) combination is set to (0.12 m, 1 s, 0 m), there are 443/213/56/55/18/11 CLoS segments lasting for 0.8/1.9/2.7/3.4/20.2/60 seconds, respectively. $P_{LoS}$ is 98.591\%, with $\mu_{CLoS}$ and $\sigma_{CLoS}$ equal to 9.61 seconds and 10.78 seconds, respectively. Moreover, when $h_{a}$ is further set to 0.1 m, or ($H_{s}$, $T_{p}$, $h_{a}$) = (0.12 m, 1 s, 0.1 m), $P_{LoS}$ is improved to 99.999\%. Statistically, there are 1/1/998 CLoS segments lasting for 3.9/22.6/60 seconds, respectively. As observed from this table, shorter $T_{p}$ or higher $H_{s}$ results in smaller values of $\mu_{CLoS}$ and $\sigma_{CLoS}$, and slightly worse $P_{LoS}$. Also, when $h_{a}$ is close to $H_{s}$, both $P_{LoS}$ and $\mu_{CLoS}$ will be significantly improved. 

\begin{table*}[h]
\small
\caption{Characterization and comparison of several combinations for maritime communication channels.} \label{tab:marine}
\newcommand{\tabincell}[2]{\begin{tabular}{@{}#1@{}}#2\end{tabular}}
 \centering
 \begin{threeparttable}
 \begin{tabular}{|c|c|c|c|c|c|c|}\hline
        \tabincell{c}{\textbf{$H_{s}$(m)}}  & \tabincell{c}{\textbf{$T_{p}$(s)}}  & \tabincell{c}{\textbf{$h_{a}$(m)}} & \tabincell{c}{\textbf{$P_{LoS}$}} & \tabincell{c}{ \textbf{$\mu_{CLoS}$(s)}} & \tabincell{c}{\textbf{$\sigma_{CLoS}$(s)}} &    
        \tabincell{c}{\textbf{$\gamma_{CLoS}$(s)}}  \\  \hline
        \tabincell{c}{0.12} &\tabincell{c}{2} & \tabincell{c}{0} & \tabincell{c}{98.595\%} & \tabincell{c}{12.98} & \tabincell{c}{13.07} & \tabincell{c}{1.1}  \\ 
        \hline
        \tabincell{c}{0.12} &\tabincell{c}{1} & \tabincell{c}{0} & \tabincell{c}{98.591\%} & \tabincell{c}{9.61} & \tabincell{c}{10.78} & \tabincell{c}{0.8}    \\ 
        \hline
        \tabincell{c}{0.12} &\tabincell{c}{1} & \tabincell{c}{0.1} & \tabincell{c}{99.999\%} & \tabincell{c}{59.88} & \tabincell{c}{2.25} & \tabincell{c}{60.0}  \\ 
        \hline
        \tabincell{c}{0.24} &\tabincell{c}{2} & \tabincell{c}{0} & \tabincell{c}{81.40\%} & \tabincell{c}{1.742} & \tabincell{c}{1.626} & \tabincell{c}{0.9}    \\ 
        \hline
        \tabincell{c}{0.67} &\tabincell{c}{2.8} & \tabincell{c}{0.4} & \tabincell{c}{95.48\%} & \tabincell{c}{7.87} & \tabincell{c}{9.16} & \tabincell{c}{0.1}   \\ 
        \hline
        \tabincell{c}{0.67} &\tabincell{c}{2.8} & \tabincell{c}{0.8} & \tabincell{c}{99.92\%} & \tabincell{c}{51.37} & \tabincell{c}{17.09} & \tabincell{c}{60.0}    \\ 
        \hline
        \tabincell{c}{2} &\tabincell{c}{9} & \tabincell{c}{1} & \tabincell{c}{99.24\%} & \tabincell{c}{41.21} & \tabincell{c}{21.07} & \tabincell{c}{60.0}    \\ 
        \hline
        \tabincell{c}{\textcolor{black}{4}} &\tabincell{c}{\textcolor{black}{10}} & \tabincell{c}{\textcolor{black}{1}} & \tabincell{c}{\textcolor{black}{84.22\%}} & \tabincell{c}{\textcolor{black}{8.78}} & \tabincell{c}{\textcolor{black}{7.88}} & \tabincell{c}{\textcolor{black}{4.6}}    \\ 
        \hline
        \tabincell{c}{\textcolor{black}{6}} &\tabincell{c}{\textcolor{black}{14}} & \tabincell{c}{\textcolor{black}{1}} & \tabincell{c}{\textcolor{black}{83.46\%}} & \tabincell{c}{\textcolor{black}{10.55}} & \tabincell{c}{\textcolor{black}{8.94}} & \tabincell{c}{\textcolor{black}{6.1}}    \\ 
        \hline
    \end{tabular}
    \end{threeparttable}
\end{table*}

\begin{figure}[t]
\begin{center}
\includegraphics[width=3.4in]{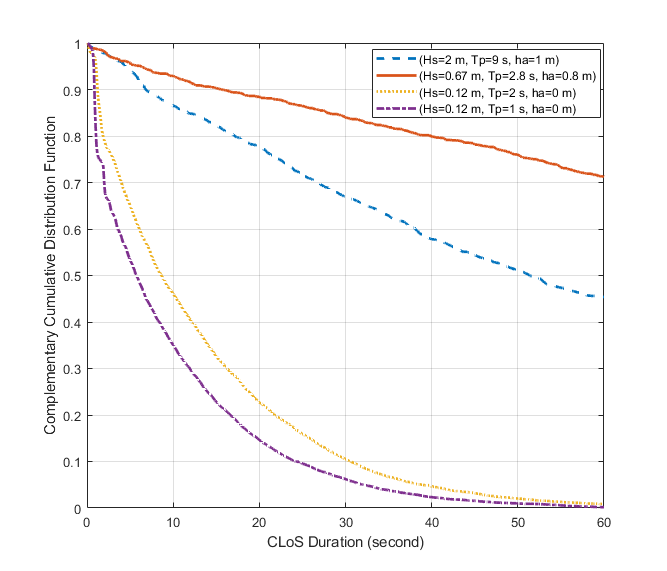}
\caption{Conditional complementary cumulative distribution function of CLoS duration for ($H_{s}$, $T_{p}$, $h_{a}$) at four combinations, respectively.}\label{fig:CCDF}
\end{center}
\end{figure}

\begin{figure}[t]
\begin{center}
\includegraphics[width=3.4in]{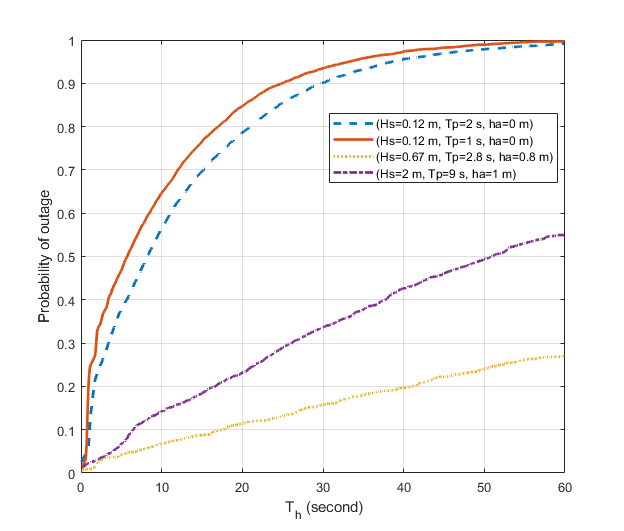}
\caption{Probability of LoS outage for ($H_{s}$, $T_{p}$, $h_{a}$) combinations at (0.12 m, 2 s, 0 m), (0.12 m, 1 s, 0 m), (0.12 m, 2 s, 0 m), (2 m, 9 s, 1 m) and (0.67 m, 2.8 s, 0.8 m) }\label{fig:Pout}
\end{center}
\end{figure}

The conditional CCDF of CLoS duration is plotted in Fig.~\ref{fig:CCDF} for four different ($H_{s}$, $T_{p}$, $h_{a}$) combinations at (2 m, 9 s, 1 m), (0.67 m, 2.8 s, 0.8 m), (0.12 m, 2 s, 0 m), and (0.12 m, 1 s, 0 m), respectively. As observed, increasing the effective antenna heights can significantly enlarge the CLoS duration. \textcolor{black}{The LoS outage probability} for four ($H_{s}$, $T_{p}$, $h_{a}$) combinations is plotted in Fig.~\ref{fig:Pout}, using \eqref{eq:Pout}. Note that these results are all based on setting the time window $T=60$ s. A larger $T$ can be used if needed. Given a desired \textcolor{black}{LoS outage probability}, $T_h$ is obtained from the curves in Fig.~\ref{fig:Pout} and from this number the data packet length can be estimated and designed to match $T_h$. Therefore, judiciously designing data packet can minimize the \textcolor{black}{LoS outage probability} and thus increase the communications performance and system energy efficiency.  

\section{Enabling Energy-Efficient Maritime IoT System Designs}

\subsection{Energy Efficiency Challenge for Maritime IoT System}

The aforementioned and analyzed features and characteristics of the maritime communications can bring along a series of challenges for adopting the cellular communications. The ocean wave blocking triggered LoS communications interrupt can result in several critical issues for the buoy’s cellular IoT communications. First, the ocean wave blockers render the loss of instant communication between the buoy and tower when the blocking criterion is satisfied. In particular, the failed uplink transmission from the buoy to the tower increases transmission retries and therefore the latency. 

Moreover, retransmission can consume significant power and degrade the system energy efficiency. On one hand, the power amplifier(s) at the transmitter end of the user equipment (UE) (buoy in this case) may need to operate, together with the buoy antenna(s), at an equivalent isotropically radiated power (EIRP) of at least 20 dBm maximum output power (23 dBm in another mode of power class). On the other hand, supporting the single-carrier frequency division multiple access (SC-FDMA) and 16 QAM modulation schemes for uplink in LTE Cat. M1 (widely used by cellular IoT) can cause a large peak to average added power ratio (PAPR) at around 6 dB \cite{Yin:PA}. By further considering the insertion loss of the RF switch(es), the cellular PAs may necessitate an extra 7-dB headroom or an equivalent 27 dBm peak power, which can further reduce the average power added efficiency (PAE) and pose more serious PA design challenges for the critical specifications such as adjacent channel leakage power ratio (ACLR) and error vector magnitude (EVM). The relationship of PA peak power consumption and UE's EIRP is 
\begin{equation}\label{eq:PA}
\begin{split}
EIRP_{UE} = P_{PA,DC}*PAE + P_{PA,IN} - IL_{RF} + G_{ANT} - \\
P_{PBO},
\end{split}
\end{equation}
where $P_{PA,DC}$ is the total DC power consumption of the standalone PA, $P_{PA,IN}$ is the input RF power, $IL_{RF}$ is the insertion loss of the RF switch, $G_{ANT}$ is the antenna gain, and $P_{PBO}$ is the PA power back-off. According to practical implementation scenarios, it is reasonable to assume, $G_{ANT}$ = 0 dBi, $P_{PBO}$  = 6 dB, $IL_{RF}$ = 1 dB, $P_{PA,IN}$ = 0 dBm. Then, considering the state-of-the-art PAE demonstrated in the \textcolor{black}{existing works, e.g., Doherty PA design \cite{Yin:PA}, CMOS PA with novel class-G efficiency enhancement for Cat. M1 and NB-IoT \cite{Bechthum:PA},} and envelope tracking (ET) technique assisted PA design \cite{Balteanu:PA}, an upper-bound PAE is set to 40\%. The total DC power consumption, $P_{PA, IN}$ is thus calculated as 1253 mW and 2500 mW, for 20 dBm and 23 dBm EIRP, respectively. Moreover, the rest 60\% of the power consumption will be transformed to, namely, power at undesired frequencies (harmonic or cross-modulated products), thermal energy, etc. Although the PA(s) system contributes a significantly large part, the total power consumption of the cellular IoT module can be much larger. The transformed thermal energy can also cause potential engineering issues in the sealed room of a compact IoT and UE device \cite{Huo:2019}.  

\textcolor{black}{Regarding the wireless system level,} the transmission interrupts and re-transmission requests will result in higher power consumption and thus a lower energy efficiency, which reduces the battery life and increases the maintenance cost for the buoy. Using the collected real-time wave data on the buoy and the analytical tools presented in this paper can assist the protocol design on adaptive data packet size and timed transmission scheduling according to the wave conditions. \textcolor{black}{In addition, latency reduction techniques, such as implementing short transmission time interval (TTI) and semi-persistent scheduling can be potentially applied to narrowband 4G LTE networks \cite{Amjad:Latency}. In order to further improve lifespan and energy efficiency of IoT devices, a new shift from legacy medium access control (MAC) to on-demand wake-up radio (WuR) operation has been considered in 3GPP Release 16 \cite{Masek:Implementation}, \cite{Anders:Wake}.}

\begin{figure}[t]
\begin{center}
\includegraphics[width=3.4in]{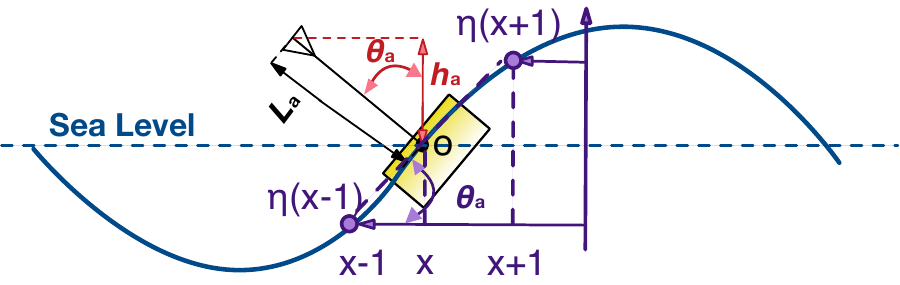}
\caption{Modeling the antenna position of a maritime IoT (buoy) device in a dynamic ocean environment.}\label{fig:Buoy_ant}
\end{center}
\end{figure}

\begin{figure}[t]
\begin{center}
\includegraphics[width=3.6in]{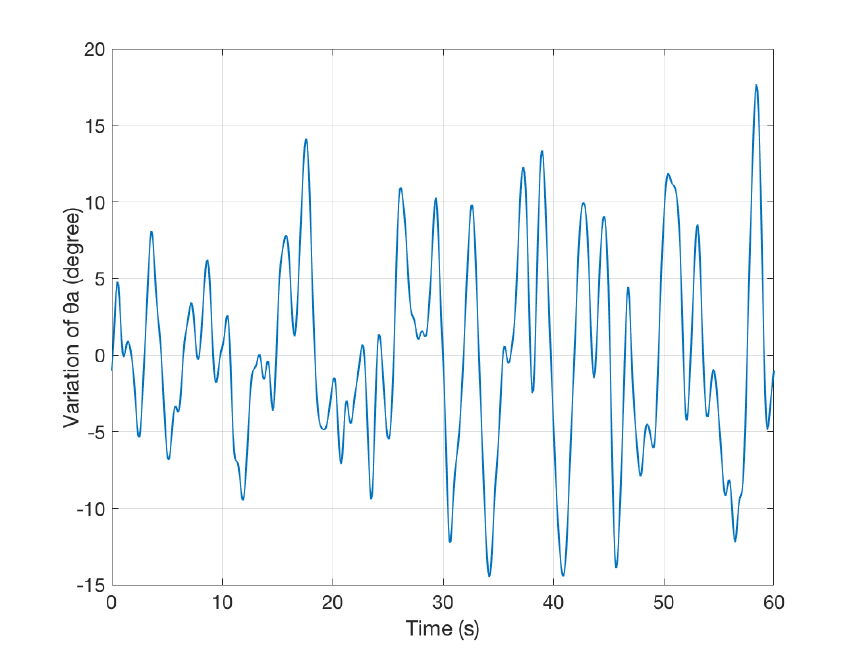}
\caption{$\theta_{a}$ variation over the time for the ocean wave combination ($H_{s}$, $T_{p}$) at (10 m, 13 s).}\label{fig:Theta}
\end{center}
\end{figure}

\begin{figure}[t]
\begin{center}
\includegraphics[width=3.6in]{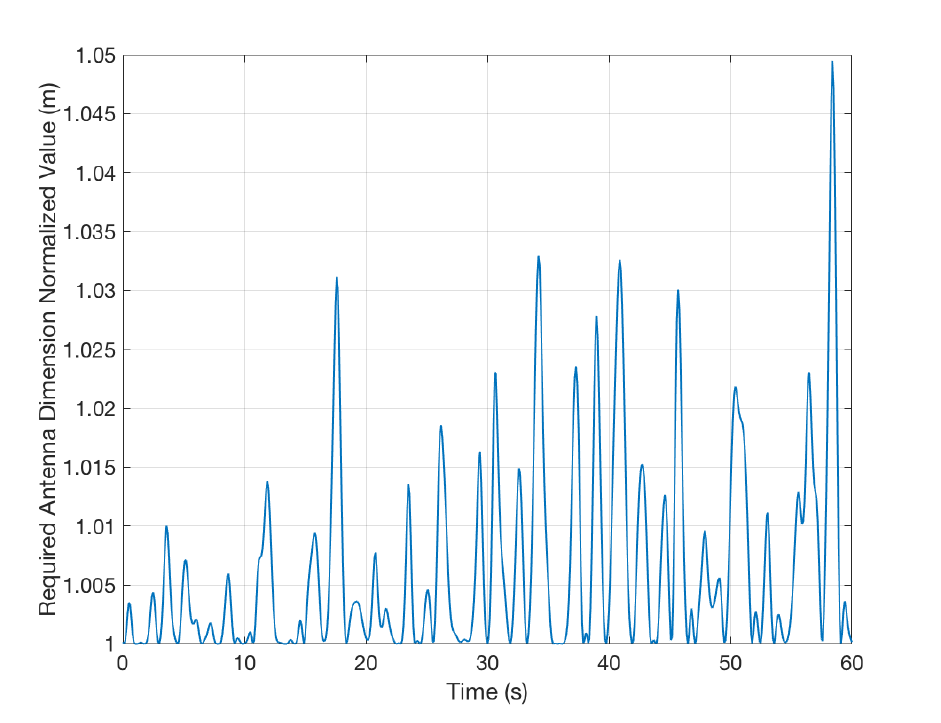}
\caption{Required (normalized) antenna dimension for the ocean wave combination ($H_{s}$, $T_{p}$) at (10 m, 13 s).}\label{fig:Theta2}
\end{center}
\end{figure}

\subsection{Antenna Hardware And System Co-design}

Based on the previous investigation and analysis, increasing the effective antenna height can significantly improve the probability of LoS communication and increase the mean values of CLoS duration. Therefore, the system energy efficiency that is directly proportional to $P_{LoS}$ will also improve when $h_{a}$ increases. On the other hand, the antenna positioning and radiation pattern affect the communication quality in a dynamic ocean environment. Hence, it is important to characterize them through hardware investigation and numerical analysis.    

First of all, installing and placing the cellular antenna outside and above the buoy’s main body is favored, although it probably leads to some mechanical and product design challenges. As drawn in Fig.~\ref{fig:Buoy_ant}, assume that the buoy (with a radius below 1 meter) constantly quasi-submerges in the ocean water with an antenna of physical dimension $L_{a}$ stand out from the point ‘O’. A time-varying angle is formed between the antenna and the sea level. Through drawing an assisted tangent line across the point ‘O’ that has horizontal coordination $x$ with an unit of meter, the angle formed between the antenna and the vertical line of sea level, $\theta_a$ (also the slope of the tangent line) can be approximated by
\begin{equation}\label{eq:theta} 
\begin{split} 
\theta_{a} = \tan^{-1}( \frac{\eta(x+1)-\eta(x-1)}{2} ), 
\end{split} 
\end{equation}
where $\eta(x)$ represents the surface elevation. The variation of $\theta_{a}$ over time can be characterized for various ocean wave conditions through a detailed numerical analysis. As illustrated in Fig.~\ref{fig:Theta}, in the extreme scenario (very rare) when the ocean wave has a very large $H_{s}$ at 10 m, the maximum $\theta_{a}$ is recorded at 17.66$^{\circ}$. Furthermore, through the relation $L_{a}$ = $h_{a}$/cos($\theta_{a}$), the impacts of the $\theta_{a}$ variation on the antenna dimension can be demonstrated.  Fig.~\ref{fig:Theta2} show the actual required normalized antenna dimension (compared to its original antenna dimension when $\theta_{a}$ is zero) over one $T$ period. Due to the time-varying $\theta_{a}$, the required $L_{a}$ also changes significantly, with a peak value of 1.049 m spotted at 58.4 second. Moreover, when the ($H_{s}$, $T_{p}$) combination is further set to (1 m, 2 s) which can be a more often observed case, a maximum $\theta_{a}$ of 9.338$^{\circ}$ is observed, which is translated to a required (normalized) antenna dimension of 1.013 m. 

\begin{figure}[!t]
    \begin{tabular}[b]{c}
       \begin{subfigure}[]{\includegraphics[width=3.9cm]{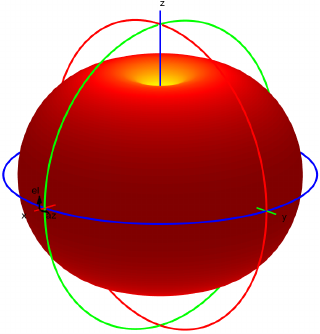} }\label{fig:A}
         \end{subfigure}
       \begin{subfigure}[]{\includegraphics[width=3.9cm]{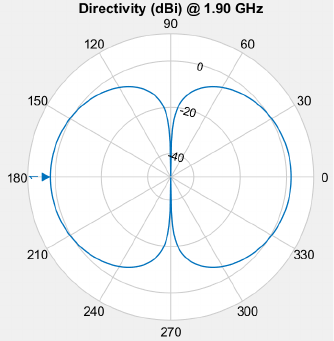}}\label{fig:B}
         \end{subfigure}
      \end{tabular}
       \begin{tabular}[b]{c}
       \begin{subfigure}[]{\includegraphics[width=9cm]{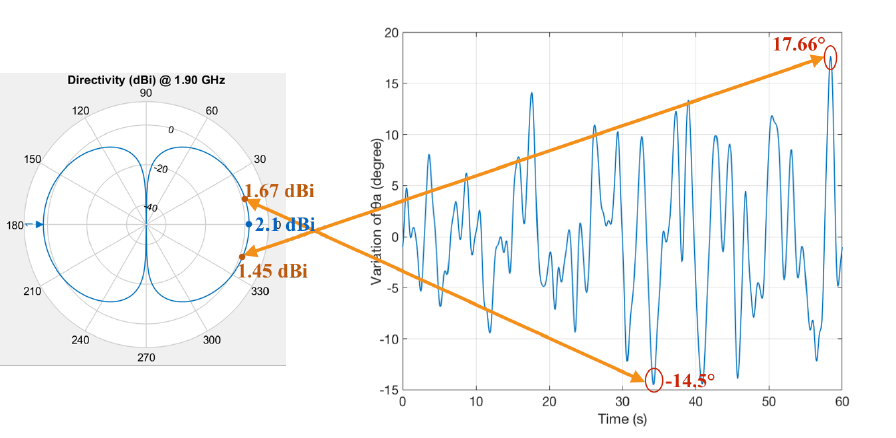}}\label{fig:C}
       \centering
       \end{subfigure}
     \end{tabular}      
     \caption{Dipole antenna (a) 3D radiation pattern, (b) elevation directivity, and (c) mapping between antenna gain and $\theta_{a}$ variation ($\Delta \theta_{a}$) when the combination ($H_{s}$, $T_{p}$) equals to (10 m, 13 s).} \label{fig:antenna}
\end{figure}

Furthermore, the $\theta_{a}$ variation needs to be mapped to the antenna and link gain variation through the antenna radiation pattern. In general, several antenna candidates that are quasi-omnidirectional can be used for buoy’s cellular communication, such as Dipole, Monopole, Bowtie, etc. Take vertically polarized dipole antenna working at 1.9 GHz center frequency (LTE Band 1 uplink) for example, its 3D radiation and elevation patterns are plotted in Fig.~\ref{fig:antenna}(a) and (b).  Fig.~\ref{fig:antenna}(c) shows the mapping between the antenna tilting angle and the radiation gain. While the maximum directivity of the antenna is 2.1 dBi, the maximum $\theta_{a}$ of 17.66$^{\circ}$ can result in a 1.45 dBi directivity. Moreover, the directivity varies between 1.45 dBi to 2.1 dBi across the entire $\theta_{a}$ variation. Several other types of antennas and combinations are investigated and summarized in Table~\ref{tab:antenna}. The listed antennas all have symmetrical quasi-omnidirectional radiation patterns. In particular, BiCone antenna has comparatively low gain with flat variation but occupies the largest bandwidth among all and can be considered for multi-band supported cellular IoT applications. It is worth mentioning that some unlisted antennas such as Yagi-Uda antenna, Vivaldi antenna, etc., have also been investigated but concluded as non-ideal candidates due to the unsymmetrical radiation pattern (only strong in one particular direction or hemisphere). 

\textcolor{black}{Finally, based on former and new (antenna and hardware system) designs have been conducted respectively, the test results and data analysis have demonstrated that increasing antenna height can combat the LoS blocking issue and significantly improve the system energy efficiency.}

\textcolor{black}{Moreover, selecting proper antenna type according to the radiation pattern and adjusting the antenna height based on a trade-off between the buoy mechanical stability and wireless performance is a key enabling factor of the entire system.}

\begin{table}[h]
\caption{Comparison of different types of antennas and directivity variation.} \label{tab:antenna}
\newcommand{\tabincell}[2]{\begin{tabular}{@{}#1@{}}#2\end{tabular}}
 \centering
 \begin{threeparttable}
 \begin{tabular}{|c|c|c|c|c|c|}\hline
        \tabincell{c}{\textbf{$H_{s}$(m)}}  & \tabincell{c}{\textbf{$T_{p}$(s)}}  & \tabincell{c}{\textbf{Max$\Delta \theta_{a}$($^{\circ}$)}} & \tabincell{c}{\textbf{Antenna} \\ \textbf{Type}} & \tabincell{c}{\textbf{Max.} \\ \textbf{$L_{a}$(m)}} & \tabincell{c}{\textbf{Directivity} \\ \textbf{(dBi)}  }   \\  \hline
        \tabincell{c}{10} &\tabincell{c}{13} & \tabincell{c}{17.66} & \tabincell{c}{Dipole} & \tabincell{c}{1.049} & \tabincell{c}{2.1-1.45}   \\ 
        \hline
        \tabincell{c}{1} &\tabincell{c}{2} & \tabincell{c}{9.338} & \tabincell{c}{Dipole} & \tabincell{c}{1.013} & \tabincell{c}{2.1-1.91}   \\ 
        \hline
        \tabincell{c}{10} &\tabincell{c}{13} & \tabincell{c}{17.66} & \tabincell{c}{Monopole} & \tabincell{c}{1.049} & \tabincell{c}{0.89-1.18}   \\ 
        \hline
        \tabincell{c}{10} &\tabincell{c}{13} & \tabincell{c}{17.66} & \tabincell{c}{Bowtie} & \tabincell{c}{1.049} & \tabincell{c}{1.43-1.94}   \\ 
        \hline
        \tabincell{c}{10} &\tabincell{c}{13} & \tabincell{c}{17.66} & \tabincell{c}{BiCone} & \tabincell{c}{1.049} & \tabincell{c}{0.56-0.66}   \\ 
        \hline
    \end{tabular}
    \end{threeparttable}
\end{table}

To summarize, the dynamic ocean environment may cause dramatic change of the antenna’s position and thus the time-varying gain variation. Since the directivity variation is directly translated into the communication link budget variation, selecting and placing proper antennas plays a crucial role in enabling more reliable maritime communications and facilitating higher system energy efficiency.

\section{CONCLUSIONS}
In this paper, we have focused on the study of maritime IoT using cellular communication technologies from both theoretical and practical perspectives. By thorough investigation and analysis of radio propagation in stochastic ocean waves, especially the wave blockage of line of sight communication link between a buoy and a land base station, we have proposed a set of analysis tools and algorithms to characterize the probability of LoS, distribution of continuous LoS duration and \textcolor{black}{LoS outage probability}, using existing ocean wave models. These tools can help design a high-performance and energy efficient maritime IoT system. Critical antenna analysis in wave conditions has led to recommendations on antenna dimension and performance prediction. Future work includes measurement campaigns on buoys to better understand the LoS and surface scattering communication channel in large ocean waves, \textcolor{black}{quantitative analysis of the communication outage probability under NLoS}, design of energy efficient protocols, maritime communication performance characterization, edge-computing for maritime IoT, machine learning aided maritime IoT systems, etc.

\end{document}